\documentclass[twocolumn,tighten]
{aastex631}

\shortauthors{Afrin, Vagnozzi \& Ghosh}
\usepackage{comment}
\usepackage{graphicx}	
\usepackage{amsmath}	
\usepackage{amsfonts}
\usepackage{float}
\usepackage{multirow}
\usepackage{textcomp}
\usepackage{wrapfig}

\usepackage{eucal}
\hypersetup{
    colorlinks=true,
    citecolor=magenta,
    linkcolor=cyan,
}
\begin{document}
\title{Tests of Loop Quantum Gravity from the Event Horizon Telescope Results of Sgr A$^*$}
\correspondingauthor{Misba Afrin}
\email{me.misba@gmail.com}
\author[0000-0001-5545-3507]{Misba Afrin}
\affiliation{Centre for Theoretical Physics,
Jamia Millia Islamia, New Delhi 110025, India}
\author[0000-0002-7614-6677]{Sunny Vagnozzi}
\affiliation{Department of Physics, University of Trento, Via Sommarive 14, 38123 Povo (TN), Italy}
\affiliation{Kavli Institute for Cosmology, University of Cambridge, Madingley Road, Cambridge CB3 0HA, United Kingdom}
\author[0000-0002-0835-3690]{Sushant G. Ghosh}
\affiliation{Centre for Theoretical Physics,
Jamia Millia Islamia, New Delhi 110025, India}
\affiliation{Astrophysics and Cosmology Research Unit, School of Mathematics, Statistics and Computer Science,\\
University of KwaZulu-Natal, Private Bag 54001, Durban 4000, South Africa}
\begin{abstract}
The Event Horizon Telescope (EHT) collaboration's image of the compact object at the galactic center is the first direct evidence of the supermassive black hole (BH) Sgr A$^*$. The shadow of Sgr A$^*$ has an angular diameter $d_{sh}= 48.7 \pm 7\,\mu$as with fractional deviation from the Schwarzschild BH shadow diameter $\delta= -0.08^{+0.09}_{-0.09}\,,-0.04^{+0.09}_{-0.10}$ (for the VLTI and Keck mass-to-distance ratios). Sgr A$^*$'s shadow size is within $~10\%$ of Kerr predictions, equipping us with yet another tool to analyze gravity in the strong-field regime, including testing loop quantum gravity (LQG). We use Sgr A$^*$'s shadow to constrain the metrics of two well-motivated LQG-inspired rotating BH (LIRBH) models characterized by an additional deviation parameter $L_q$, which recover the Kerr spacetime in the absence of quantum effects ($L_q \to 0$). When increasing the quantum effects through $L_q$, the shadow size increases monotonically, while the shape gets more distorted, allowing us to constrain the fundamental parameter $L_q$. We use the astrophysical observables shadow area $A$ and oblateness $D$ to estimate the BH parameters. It may be useful in extracting additional information about LIRBHs. While the EHT observational results completely rule out the wormhole region in the LIRBH-2, a substantial parameter region of the generic BHs in both models agrees with the EHT results. We find that the upper bounds on $L_q$ obtained from the shadow of Sgr A$^*$ ---$L_q \lesssim 0.0423$ and $L_q \lesssim 0.0821$ for the two LIRBHs, respectively---are more stringent than those obtained from the EHT image of M87$^*$.
\end{abstract}
\keywords{Astrophysical black holes (98); Black hole physics (159); Galactic center (565);  Gravitation (661); Gravitational lensing (670)}
\section{Introduction}\label{Intro}
\begin{figure*}[t]
\begin{center}
    \begin{tabular}{c c}
   \hspace{-0.6cm} \includegraphics[scale=0.80]{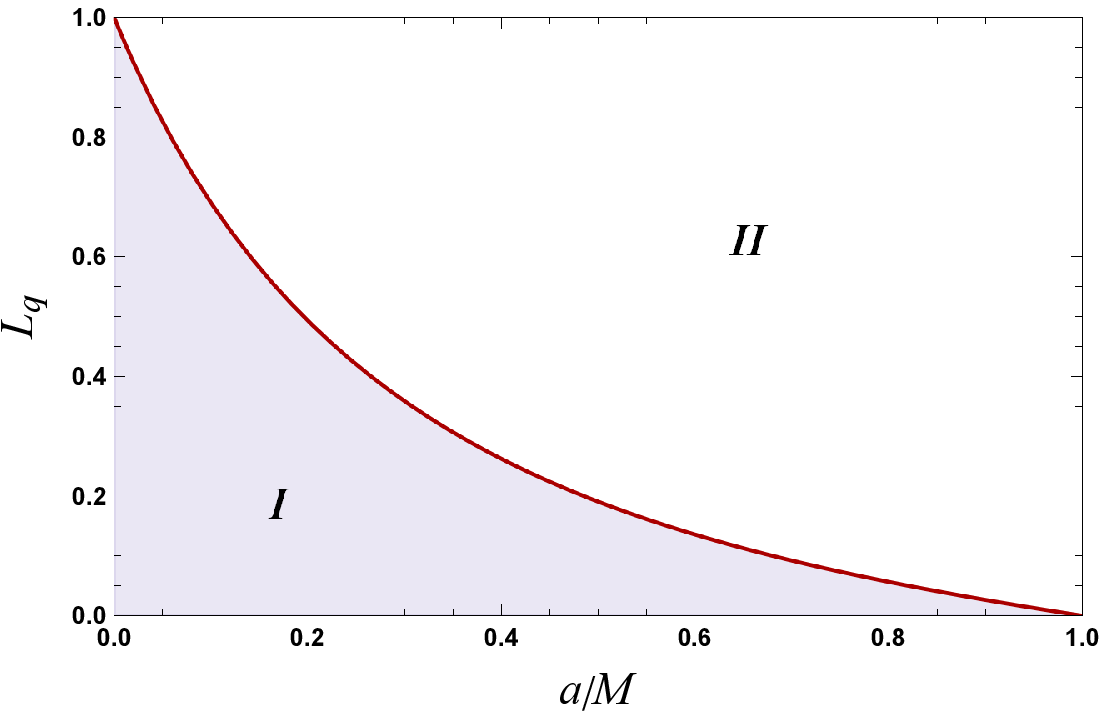}&
     \includegraphics[scale=0.80]{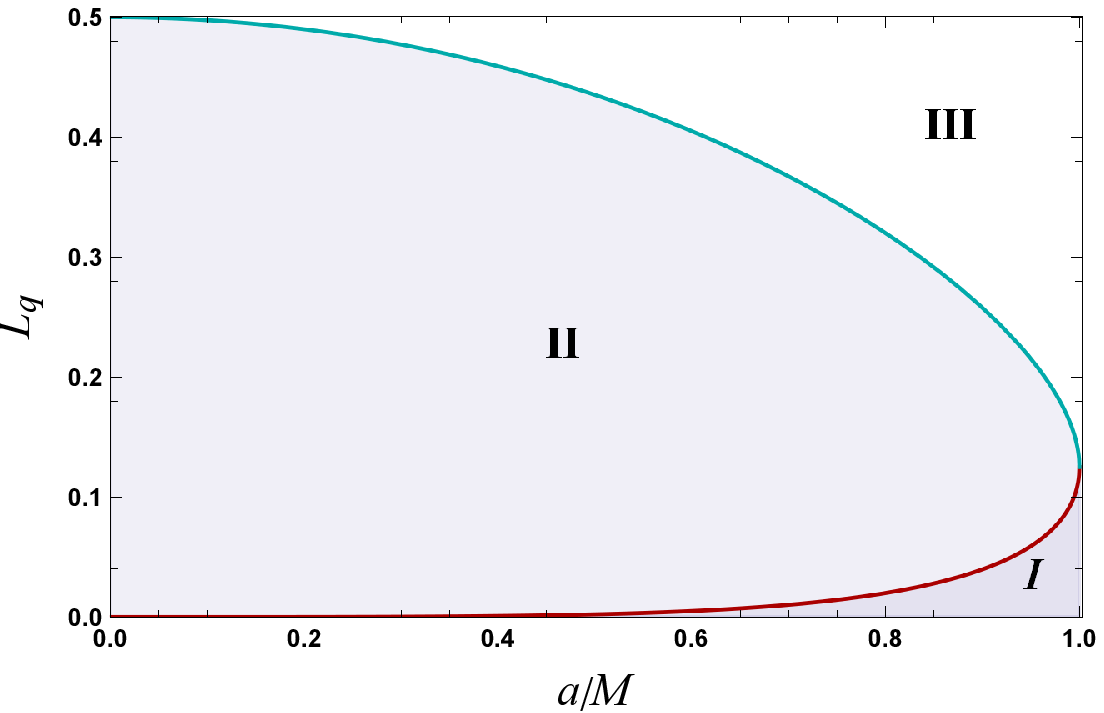}
\end{tabular}
\end{center}
	\caption{Left: parameter space of LIRBH-1. Region I represents generic BHs with Cauchy and event horizons, while the region II is a no-horizon spacetime \citep{Liu:2020ola}. Right: parameter space of LIRBH-2. Region I represents generic BHs with Cauchy and event horizons, region II is a BH with a single horizon, and region III is a wormhole, which we shall show to be ruled out by EHT observations \citep{Brahma:2020eos}.}
	\label{fig:parameterspace}
\end{figure*}
The theory of general relativity (GR), though being a widely tested standard model of gravity with remarkable consequences, such as the existence of black holes \citep[BHs;][]{Schwarzschild:1916uq}and gravitational waves, is nevertheless not free of pathologies, which has led to calles for modifications and alternatives to it~\citep{Nojiri:2017ncd}. The necessary extension of GR at the quantum scale was emphasized by Einstein himself \citep{Einstein:1916cc}, and thus far there have been several efforts in this direction \citep[see][for a recent review]{Addazi:2021xuf}, with the most promising candidate quantum gravity models being provided by string theory and loop quantum gravity~\citep[LQG; see][for a review]{Rovelli:1997yv}.

BHs are among the many fascinating objects in the Universe, enveloped by matter under extreme conditions in a regime of strong spacetime curvature. Studying these BHs can lead to a greater insight into their nature, circumstances, and significance for fundamental theories like GR and other theories like LQG. The no-hair theorem says that three parameters describe isolated and stationary BHs in GR, namely mass $M$, spin $J$, and electric charge $Q$. Thus, they are defined by the Kerr-Newman metric \citep{Newman:1965my}, which goes to the Kerr metric \citep{Kerr:1963ud} that are charge neutral. The Kerr BH describes the astrophysical BHs because any residual electric charge is expected to rapidly neutralize \citep{Israel:1967wq,Israel:1967za,Carter:1971zc,Carter:1999mrq,Hawking:1971vc}. Indeed, direct evidence of BH charge neutrality is still inconclusive, and it may be difficult to rule out non-Kerr BHs \citep{Ryan:1995wh,Will:2005va}. In addition, the no-hair theorem's mathematical status is not without controversy, principally concerning the assumption of analyticity \citep{Chrusciel:2012jk}. One can test the no-hair theorem by calculating potential deviations from Kerr metrics like the LQG-motivated rotating BHs. The celebrated theorem does not hold for the modified theories of gravity, like LQG, that admit non-Kerr BHs. The images of the supermassive BHs 
M87* \citep{EventHorizonTelescope:2019dse,EventHorizonTelescope:2019pgp,EventHorizonTelescope:2019ggy} and Sgr A* \citep{EventHorizonTelescope:2022exc,EventHorizonTelescope:2022xqj} released by the Event Horizon Telescope (EHT) collaboration led us into an untouched stage of BH physics, offering a direct visualization of supermassive BHs, as well as their surrounding environment.
Utilizing the distance of M87* from Earth of $16.8$ Mpc and the estimated mass of $(6.5 \pm 0.7) \times 10^9 M_\odot$, puts bounds on the compact emission region size with angular diameter $d_{sh}=42\pm 3\, \mu $as and circularity deviation $\Delta C\leq 0.10$. Recently, the EHT collaboration reported the Sgr A* BH shadow results \citep{EventHorizonTelescope:2022exc,EventHorizonTelescope:2022urf,EventHorizonTelescope:2022vjs,EventHorizonTelescope:2022wok,EventHorizonTelescope:2022xnr,EventHorizonTelescope:2022xqj}; considering a BH of mass $4.0^{+1.1}_{-0.6} \times 10^6 M_\odot $ and distance $8$kpc from Earth, the EHT results reveal that the shadow of Sgr A* has an angular diameter $d_{sh}= 48.7 \pm 7\,\mu$as with fractional deviation from the Schwarzschild BH shadow diameter $\delta= -0.08^{+0.09}_{-0.09}\,,-0.04^{+0.09}_{-0.10}$ (for the VLTI and Keck mass-to-distance ratios, respectively), and the images are consistent with the expected appearance of a Kerr BH \citep{EventHorizonTelescope:2022exc,EventHorizonTelescope:2022xqj}. Compared with the EHT results for M87*, it reveals consistency with the predictions of GR \citep{EventHorizonTelescope:2022xnr}.

The EHT observation of M87* and Sgr A* presents a new powerful technique to test the BH metric gravitationally in the strong-field regime and also provides an exceptional way to constrain the various BH parameters and to test the underlying associated theories of gravity \citep{EventHorizonTelescope:2021dqv,EventHorizonTelescope:2022xqj,Ghosh:2020spb,Afrin:2021imp,KumarWalia:2022aop,Kumar:2022fqo,Islam:2022ybr,Sengo:2022jif}. Therefore, the EHT results offer a considerable recent complement to the set of observations that probe the strong-field regime of gravity, namely the LIGO/Virgo detection of gravitational waves from stellar-mass BH mergers \citep{LIGOScientific:2016aoc}.  
The observations of supermassive BHs M87* and Sgr A* can offer more compelling prospects for testing LQG. LQG, being a nonperturbative approach to quantum gravity, goes beyond GR to resolve classical spacetime singularities in the BH spacetimes \citep{Ashtekar:2006wn,Ashtekar:2006es,Vandersloot:2006ws}. Because of the inherent hardship in solving the complete system, the emphasis has been on spherically symmetric BH spacetimes. In the semiclassical regimes, within the framework of LQG, it turns out that several spherically symmetric BH models exist such that singularity occurring in the GR is now substituted by a transition regular surface \citep{Ashtekar:2005qt,Boehmer:2007ket,Modesto:2008im,Perez:2017cmj,Gambini:2008dy,
Gambini:2013ooa,Corichi:2015xia,Olmedo:2017lvt,Ashtekar:2018lag,Ashtekar:2018cay,Bodendorfer:2019xbp,Bodendorfer:2019cyv,Arruga:2019kyd,Assanioussi:2019twp,BenAchour:2020gon,Gambini:2020nsf,Bodendorfer:2019nvy,Bodendorfer:2019jay,Blanchette:2020kkk,Assanioussi:2020ezr,Chen:2022nix}.

We obtain LQG-inspired rotating BHs (LIRBHs) via the revised Newman-Janis generating method \citep{Liu:2020ola,Brahma:2020eos}, which works quite well in generating rotating metrics starting with their nonrotating seed metrics arising in the modified gravities, including LQG \citep{Azreg-Ainou:2014pra,Brahma:2020eos,Liu:2020ola,Chen:2022nix,Modesto:2008im}. The LIRBHs or Kerr-like BHs, which have an additional parameter ($L_q$) coming from the quantum effects, apart from mass ($M$) and rotation parameter ($a$), can be appropriately tested with astrophysical observations. We also show that it is possible, in principle, to constrain the LQG parameter $L_q$ using the EHT observed shadow image cast by M87* and Sgr A*. Further, we aim to investigate whether the shadow images of Sgr A* can help us better determine whether the two Kerr-like LQG-motivated BHs can be suitable candidates for astrophysical BHs.  We also examine whether the EHT bounds for Sgr A* can provide more stringent constraints on the LQG BH parameter than previously acquired with the bounds for M87* observations \citep{Brahma:2020eos}. 

We organize the paper as follows: In Section~\ref{sec:methodology}, we study the effect of the parameter $L_q$  on the photon geodesics and shadow silhouettes of LIRBHs. Section~\ref{sect:estimation} is dedicated to the estimating the BH parameters $L_q$ and $a$ utilizing the shadow observables $A$ and $D$. In Section~\ref{sect:constraining}, we constrain $L_q$ from the EHT-deduced bound on the shadow observables $d_{sh}$ and $\delta$. Finally, we conclude in Section~\ref{sect:conclusion}.

Throughout this paper, we work with geometrized units, $8 \pi G=c=1$,  unless units are specifically defined.
\section{Methodological framework}\label{sec:methodology}
Here, we examine two well-motivated models, namely LIRBH-1\citep{Modesto:2008im,Liu:2020ola} and LIRBH-2 \citep{Brahma:2020eos,Yang:2022yvq}. 
 The shadows of these two models have received considerable attention, and their shape and size are substantially different from those of the Kerr BH shadows \citep{Brahma:2020eos,Yang:2022yvq,Liu:2020ola,Devi:2021ctm,KumarWalia:2022ddq}. 
The LIRBHs in question belong to a family of prototype non-Kerr BH metrics with an additional deviation parameter $L_q$ related to the quantum effects, besides $a$ and $M$ of Kerr BH. They include the Kerr BH as a particular case of vanishing quantum effects $L_q\to0$, 
and both LIRBHs provide singularity resolution of Kerr BHs.
 The BH shadow \citep{Bardeen:1973tla,Luminet:1979nyg}, a purely geometry-dependent strong-field construct can, in principle be used to determine the properties of the BH spacetime, e.g., computation of parameters \citep{Kumar:2018ple}.
 Hence, we examine the shadows of LIRBHs \citep{Liu:2020ola,Brahma:2020eos}, whose line element in Boyer$-$Lindquist coordinates ($t, r, \theta, \phi$) can be cast in a Kerr-like form \citep{Azreg-Ainou:2014pra}
\begin{eqnarray}\label{metric}
ds^2 & = & -\frac{\Psi}{\rho^2}\Bigg[\frac{\Delta}{\rho^2}(dt-a \sin^2 \theta d\phi^2)^2-\frac{\rho^2}{\Delta} dr^2   \nonumber\\
&&-\rho^2 d \theta^2 -\frac{\sin^2 \theta}{\rho^2}\big[a dt-(\omega(r)+a^2)d \phi\big]^2 \Bigg],
\end{eqnarray}
where $\rho^2 =\omega(r) + a^2 \cos^2\theta$, and $\omega(r)$ and $\Delta(r)$ are model-dependent metric functions; the Kerr-like form of metric (\ref{metric}) brings out the spacetime symmetries and simplifies solving the geodesic equations of motion significantly. 

\paragraph{LIRBH-1}
This is the rotating counterpart \citep{Liu:2020ola} of the semiclassical LQG-inspired spherical  solution \citep{Modesto:2008im}. The  LIRBH-1 is described by metric (\ref{metric}) \citep{Liu:2020ola} with 
	\begin{eqnarray}
		\Delta(r)&=& \frac{(r-r_+)(r-r_-)r^2}{(r+r_*)^2}+a^2,\\
	\omega &=& \frac{r^4+a_0^2}{(r+r_*)^2},\;\;
    \Psi(r)= \frac{r^4+a_0^2}{r^2}.
	\end{eqnarray}
Here $r_+=2 M/(1+L_q)^2$, $r_{-} = 2 M L_q^2/(1+L_q)^2$ and $r_{*}= \sqrt{r_+ r_-} = 2 ML_q/(1+L_q)^2$; $L_q=(\sqrt{1+\gamma^2 \delta^2}-1)/(\sqrt{1+\gamma^2 \delta^2}+1)$ is the polymeric function, where $\gamma$ is the Immirzi parameter  and $\delta$ is the polymeric parameter such that $\gamma \delta \ll 1$. The  ADM mass and BH spin are denoted by $M$ and $a$, respectively. In addition, the parameter $a_0 = \mathcal{A}_{\rm min}/8\pi$ is related to the minimum area gap of LQG, $ \mathcal{A}_{\rm min}=8\pi \ell_P^2 \gamma \sqrt{s_m(s_m+1)}$, where $\ell_P$ is the Planck length and $s_m$ is the smallest value of the representation on the edge of the spin network crossing the surface \citep{Santos:2015gja}; considering SU(2) group representation, we have $s_m=1/2$, and we further set $\gamma\sim1$ ~\footnote{We note that theoretical arguments lead us to expect that the Barbero--Immirzi parameter should be of order $0.1$ (e.g., either $\gamma=\ln 2/\sqrt{3}\pi$ or $\gamma=\ln 3/2\sqrt{2}\pi$), which is not definite, and the consensus on the value has changed a couple of times \citep{Modesto:2009ve}. However, we have explicitly verified that the choice of $\gamma$ has virtually no effect on our results as far as the BH shadow is concerned, as already noted earlier in~\cite{Liu:2020ola} and \cite{Vagnozzi:2022moj}. Hence, here we will stick to $\gamma \sim 1$ for simplicity.}, and thus $a_0=\sqrt{3}/2 \ell_P^2$ \citep{Santos:2015gja}. Further, in the limit $L_q=0=a_0$, the metric (\ref{metric}) goes over to the Kerr solution. A root analysis of $\Delta={(r-r_+)(r-r_-)r^2}/{(r+r_*)^2}+a^2=0$, yields a parameter space ($L_q$, $a/M$) where two real roots corresponding to the  horizons are obtained (region I in the left panel of Figure~\ref{fig:parameterspace}) as well as a parameter space corresponding to no-horizon regular spacetime where no real roots of $\Delta$ are obtained (region II in the left panel of Figure~\ref{fig:parameterspace}). Shall test the LIRBH-1 model with the EHT results of M87* and Sgr A*, which still needs to be done.
\paragraph{LIRBH-2}
This is derived using the modified Newman–Janis algorithm \citep{Brahma:2020eos,Yang:2022yvq} with the spherical LQG-inspired quantum extension of Schwarzschild spacetime \citep{Bodendorfer:2019nvy} as a seed metric. Again, the LIRBH-2 is described by the metric (\ref{metric}) with metric functions
\begin{align}
 \Delta(r) & =8L_q M_B^2\tilde{a}b^2+a^2,  \\
 \omega & =b^2,\;\;
    \Psi(r)= \rho^2
\end{align}
with
\begin{eqnarray}
   b^2(x) &=&\frac{L_q}{\sqrt{1+x^2}} \frac{M_B^2(x+\sqrt{1+x^2})^6+M_B^2}{(x+\sqrt{1+x^2})^3}, \label{bx2}\\
   \tilde{a}(x)  & =&\left(1-\frac{1}{\sqrt{2L_q}}\frac{1}{\sqrt{1+x^2}}\right)\frac{1+x^2}{b(x)^2}.
\end{eqnarray}
Here $x=r/(\sqrt{8L_q}M_B)\in(-\infty, \infty)$; $L_q=(l_k/M_B M_W)^{2/3}/2 \geq 0 $ is a dimensionless parameter, where the quantum parameter $l_k$ arises from holonomy modifications~\citep{Bodendorfer:2019nvy,Bodendorfer:2019jay} and it is directly related to the minimum area gap of the LQG theory with the areal radius given by Eq.~(\ref{bx2}) ~\citep{Brahma:2020eos}. Here, $M_B$ and $M_W$ correspond to the mass of black hole and white hole \citep{Bodendorfer:2019nvy,Bodendorfer:2019jay}, respectively; we are interested in the case $M_B=M_W=M$, i.e., a symmetric bounce. Again, solving $\Delta=0$ for real roots segregates the ($L_q$, $a/M$) parameter space into three regions as shown in Figure~\ref{fig:parameterspace}: (i) region I, is a generic BH  with two horizons, which is the most physically relevant region (ii); region II, denoting a BH with a single event horizon; and (iii) region III, corresponding to a horizon-less wormhole. We are interested only in BHs, i.e., regions I and II, as observational results of M87* have ruled out the wormhole region \citep{Brahma:2020eos}. In addition, we intend to constrain the parameter $L_q$  with observations of Sgr A* and estimate $L_q$ using the BH shadow.
\subsection{Shadow silhouette}
The EHT images of supermassive BHs exhibit a dark brightness depression surrounded by a bright ringlike feature \citep{EventHorizonTelescope:2019dse,EventHorizonTelescope:2022xnr}. The image is composed of an emission component that is mainly determined by the relatively uncertain radiative and accretion physics, which -- despite morphologically depending on the spacetime structure \citep{Ricarte:2022wpd,Palumbo:2020flt} -- is expected to be only subdominantly affected by the underlying modified theory of gravity \citep{Johannsen:2015hib}, at the current resolution of the EHT. In addition, there is a series of bright rings asymptotically spiraling and approaching the dark region \citep{Bardeen:1973tla,1973ApJ...183..237C,Luminet:1979nyg}, which is almost solely determined by the background metric alone \citep{Johannsen:2015hib,Broderick:2022tfu}. This boundary, the BH shadow silhouette, though not fully resolved yet by the EHT, can be obtained analytically as the locus of gravitationally lensed photons traveling in close proximity of the BH, onto the observer's celestial plane. Besides, it is independent of the various astrophysical phenomena \citep{Johnson:2019ljv}. 
We can employ the BH shadow as a tool to test modified theories of gravity, besides constraining the potential deviations from the Kerr metric; it has thus eventuated in a comprehensive literature addressing shadows in both GR \citep{Vries2000TheAS,Falcke:1999pj,Shen:2005cw,Yumoto:2012kz,Atamurotov:2013sca,Abdujabbarov:2015xqa,Johannsen:2015hib,Cunha:2018acu,Kumar:2018ple,Afrin:2021ggx,Chael:2021rjo} and modified theories of gravity \citep{Amarilla:2010zq,Johannsen:2010gy,Amir:2017slq,Singh:2017vfr,Kumar:2017tdw,Mizuno:2018lxz,Allahyari:2019jqz,Papnoi:2014aaa,Kumar:2019ohr,Kumar:2020bqf,Kumar:2020hgm,Kumar:2020owy,Brahma:2020eos,Ghosh:2020spb,Afrin:2021wlj,Vagnozzi:2022moj,Vagnozzi:2019apd,Afrin:2021imp,KumarWalia:2022aop,Kumar:2022fqo,Islam:2022ybr,Sengo:2022jif,Kuang:2022ojj,Junior:2021svb}. The fact that the silhouette of BH shadows encode in them, the strong-field properties of the spacetime, suggests that, we can use them for performing strong-field gravitational tests \citep{Johannsen:2010ru,Cunha:2018acu,Baker:2014zba}.
The lightlike geodesics in the LQG spacetime (\ref{metric}), just as in the Kerr spacetime, follow the Hamilton--Jacobi equation \citep{Carter:1968rr}, 
\begin{eqnarray}
\label{HmaJam}
\frac{\partial \mathcal{S}}{\partial \lambda} = -\frac{1}{2}g^{\alpha\beta}\frac{\partial \mathcal{S}}{\partial x^\alpha}\frac{\partial \mathcal{S}}{\partial x^\beta},
\end{eqnarray}
where $\lambda$ is the affine parameter along the geodesics and $\mathcal{S}$ is the Jacobi action given by
\begin{eqnarray}
\mathcal{S}=-{\mathcal E} t +{\mathcal L}_z \phi + \mathcal{S}_r(r)+ \mathcal{S}_\theta(\theta) \label{action},
\end{eqnarray}
where the conserved photon energy $\mathcal{E}=-p . \partial_{t}$ and axial angular momentum $\mathcal{L}_z=p . \partial_{\phi}$ arise as a result of the translational and rotational symmetry of the metric (\ref{metric}). Further,
the Kerr-like BHs possesses (\ref{metric}) a fourth conserved quantity, the Carter constant $\mathcal{Q}$ \citep{Brahma:2020eos,Liu:2020ola}, which ensures the decoupling of the$r$ and $\theta$ equations \citep{Carter:1968rr}. Following \cite{Tsukamoto:2017fxq,Kumar:2018ple,Liu:2020ola,Brahma:2020eos,Kumar:2020ltt}, we find that the Hamilton-Jacobi equations are separable, and we obtain null geodesics in the first-order differential form 
\begin{align}
\rho^2 \frac{dt}{d\lambda}=&\frac{\omega(r)+a^2}{\Delta}[\mathcal{E}(\omega(r)+a^2)-a\mathcal{L}_z]-a(a\mathcal{E}\sin^2{\theta}-\mathcal{L}_z),\label{TimeEq}\\
\rho^2 \frac{dr}{d\lambda}=&\pm\sqrt{\Re(r)}\ ,\label{RadialEq} \\
\rho^2 \frac{d\theta}{d\lambda}=&\pm\sqrt{\Theta(\theta)}\ ,\label{Theta}\\
\rho^2 \frac{d\phi}{d\lambda}=&\frac{a}{\Delta}[\mathcal{E}(\omega(r)+a^2)-a\mathcal{L}_z]-\left(a\mathcal{E}-\frac{\mathcal{L}_z}{\sin^2{\theta}}\right)\label{PhiEq},
\end{align}
where ${R}(r)$ and $\Theta(\theta)$ refer to radial and polar effective potentials, respectively, as follows:
\begin{eqnarray}\label{06}
\mathcal{R}(r)&=&\left((\omega(r)+a^2){\mathcal{E}}-a{\mathcal {L}_z}\right)^2-\Delta({\cal K}+(a{\mathcal{E}}-{\mathcal {L}_z})^2)\label{R_potential},\quad \\ 
\Theta(\theta)&=&{\cal K}-\left(\frac{{\mathcal {L}_z}^2}{\sin^2\theta}-a^2 {\mathcal{E}}^2\right)\cos^2\theta.\label{theta_potential}
\end{eqnarray}
Here the separability constant $\mathcal{K}=\mathcal{Q}-(a\mathcal{E}-\mathcal{L}_z)^2$ is related to the nonapparent symmetries of metric (\ref{metric}) through a quadratic Killing tensor \citep{Hioki:2008zw}. Interestingly, Eqs.~(\ref{TimeEq})-(\ref{theta_potential}) have the same mathematical form as in the Kerr case and, further, for $L_q=0$ reduce exactly to those of the Kerr case \citep{Chandrasekhar:1985kt}. The $\mathcal{Q}$ is related to the $\theta$-velocity of the photon, and for $\mathcal{Q}=0$ the photon motion is restricted to the equatorial plane, while the $\mathcal{L}_z$ controls the $\phi$-motion \citep{Chandrasekhar:1985kt,Teo:2020sey}.
\begin{figure*}[t]
\centering
\begin{tabular}{c c}
    \hspace{-1cm}\includegraphics[scale=0.7]{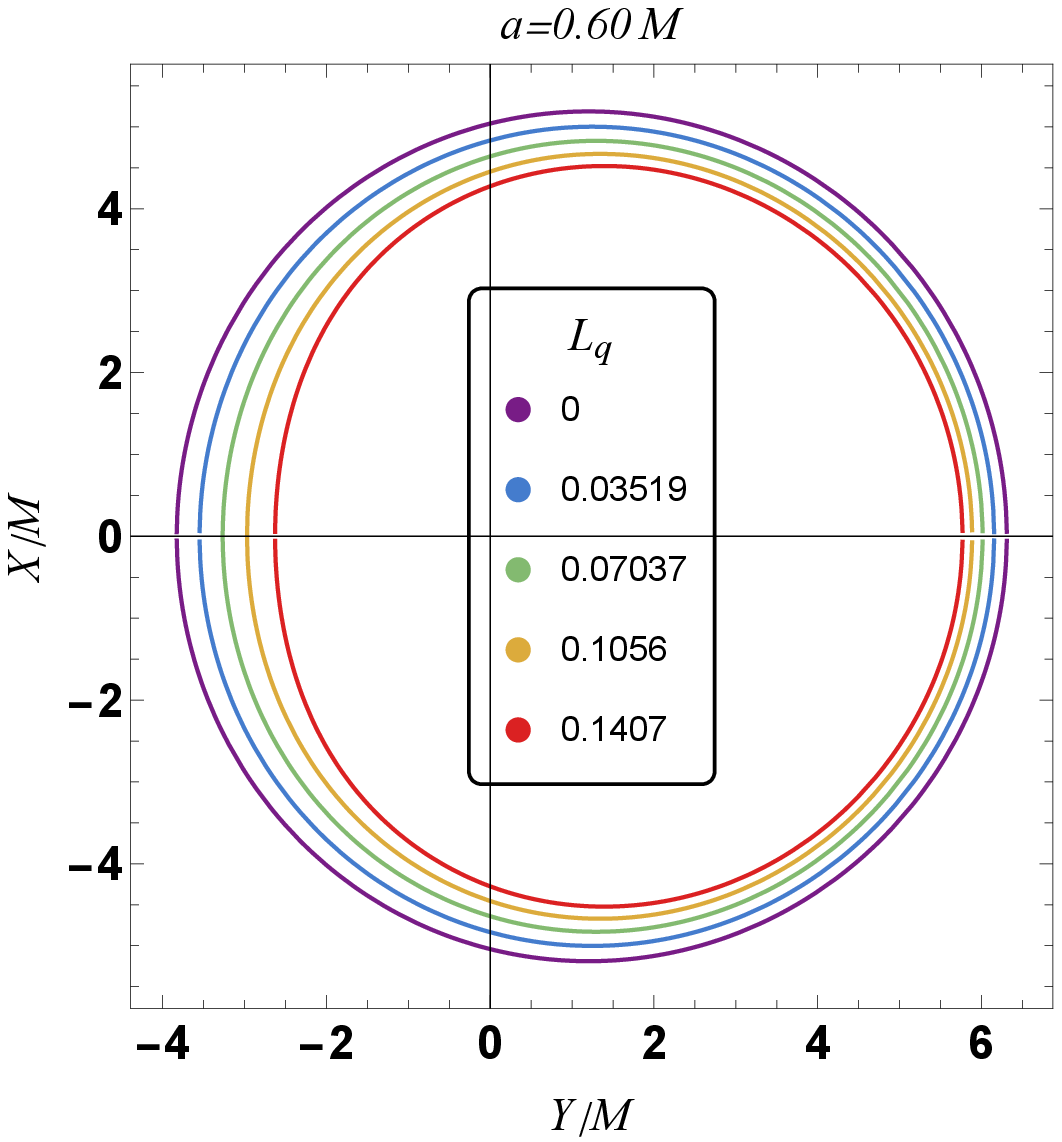}&
    \includegraphics[scale=0.67]{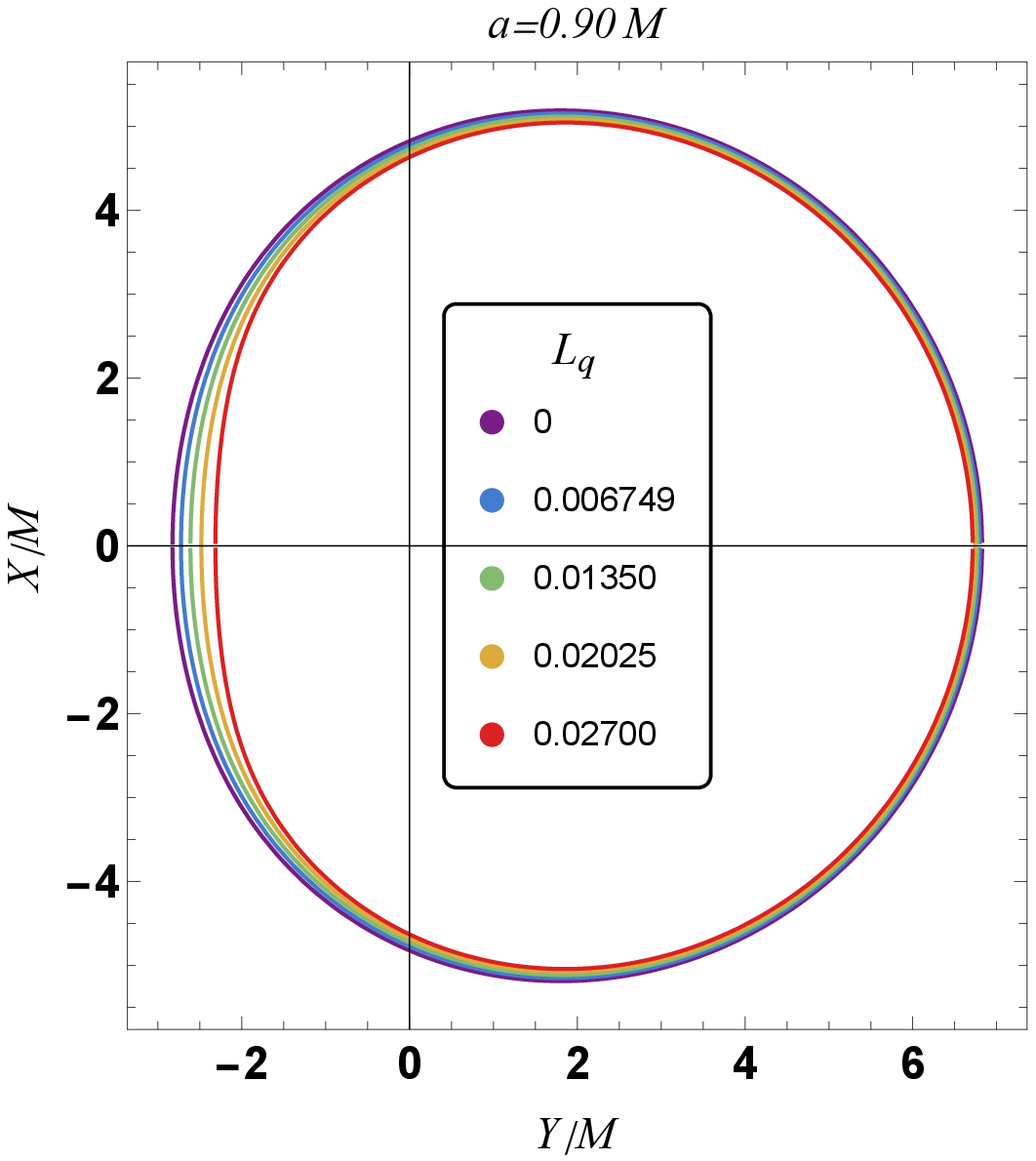}\\ 
    \hspace{-1cm}\includegraphics[scale=0.7]{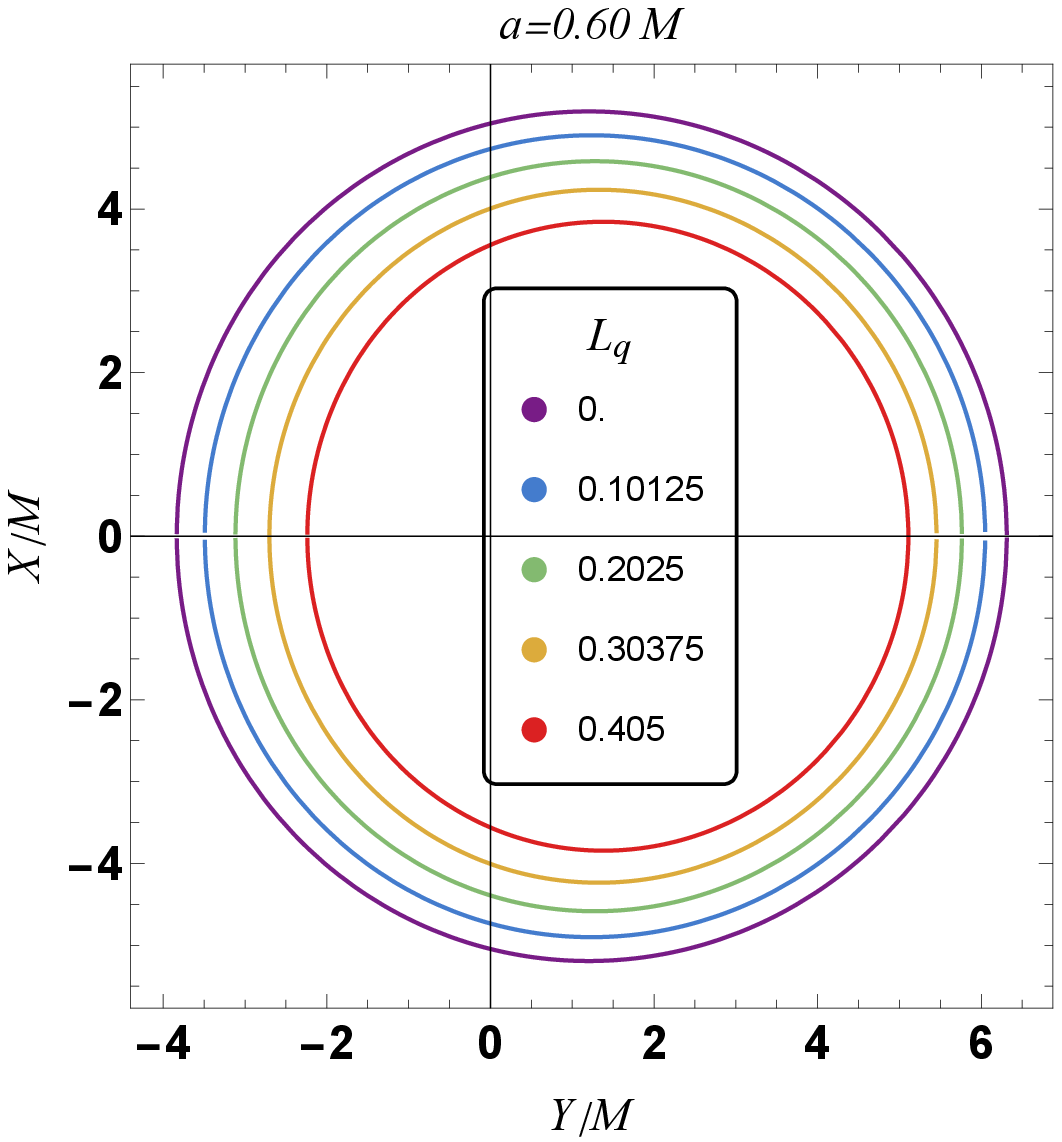}&
    \includegraphics[scale=0.67]{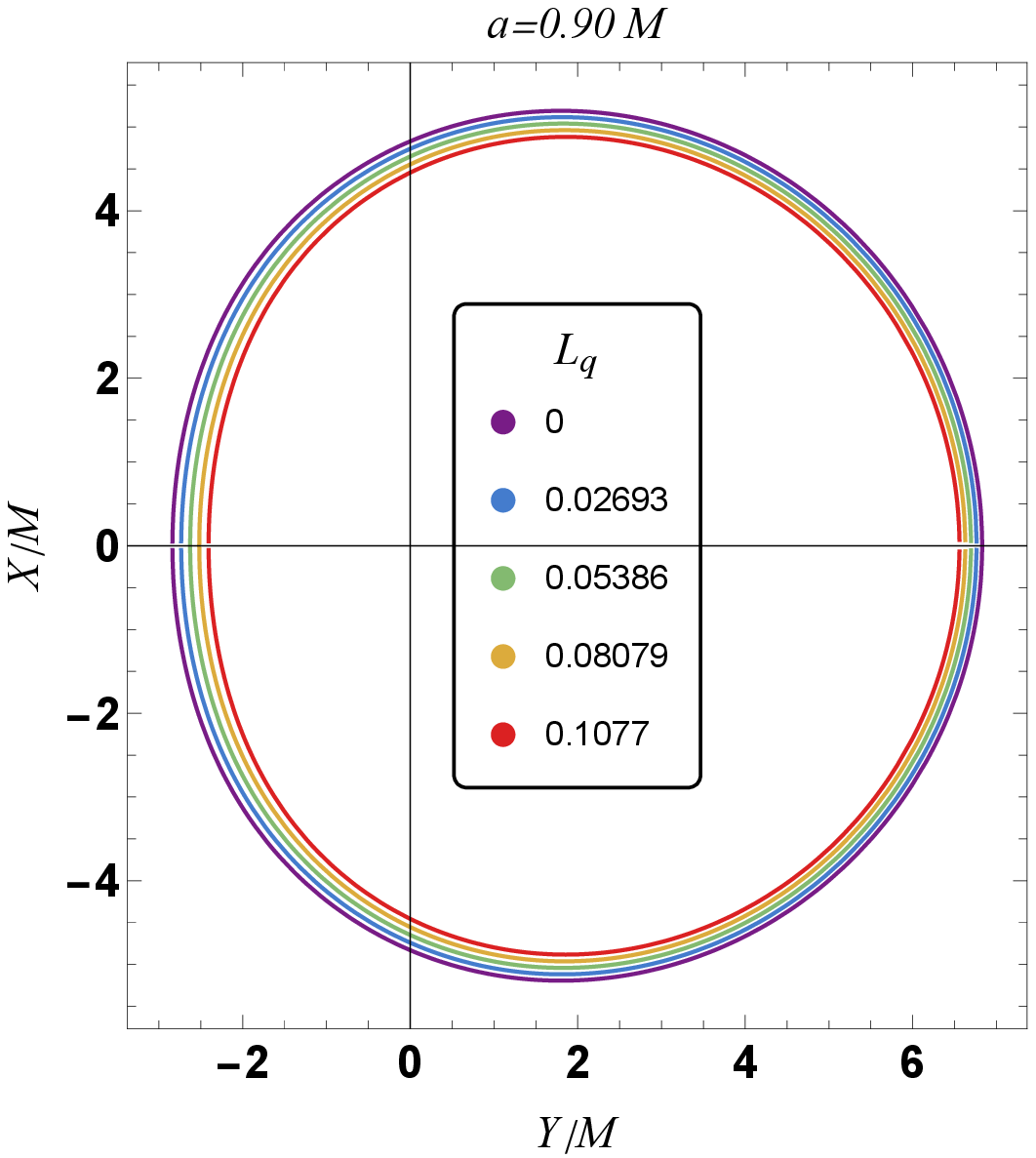}
\end{tabular}
\caption{Shadow silhouettes cast by LIRBH-1 (top) and LIRBH-2 (bottom) with varying parameter $L_q$. The solid violet curves correspond to Kerr BH ($L_q =0$) shadows.}\label{shadow_Figure}
\end{figure*}
\begin{figure*}[t]
\begin{center}
    \begin{tabular}{c c}
   \hspace{-0.7cm} \includegraphics[scale=1.05]{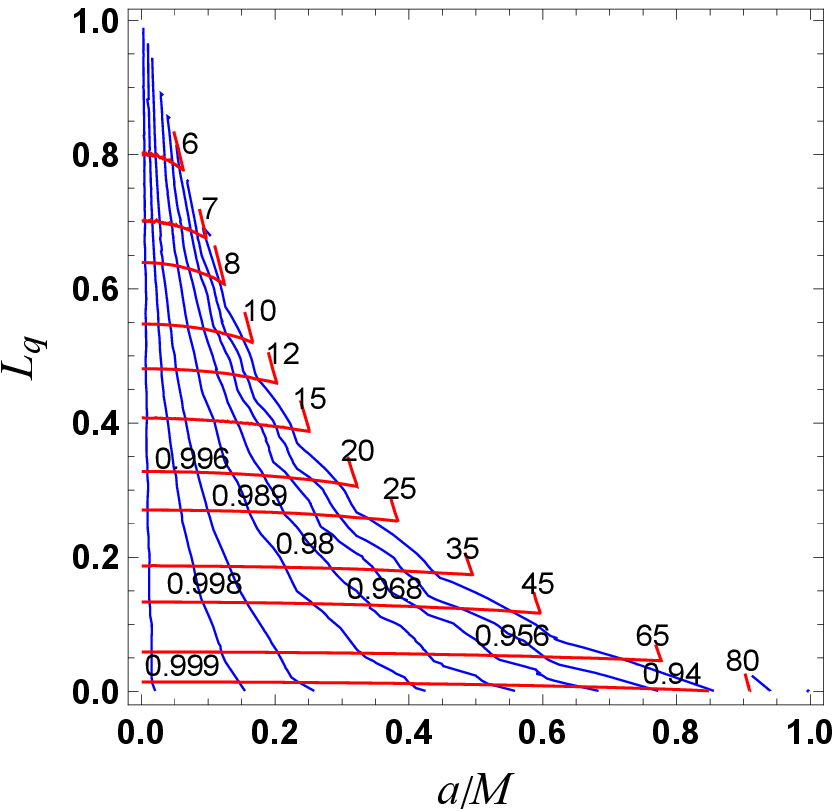}&
     \includegraphics[scale=1.05]{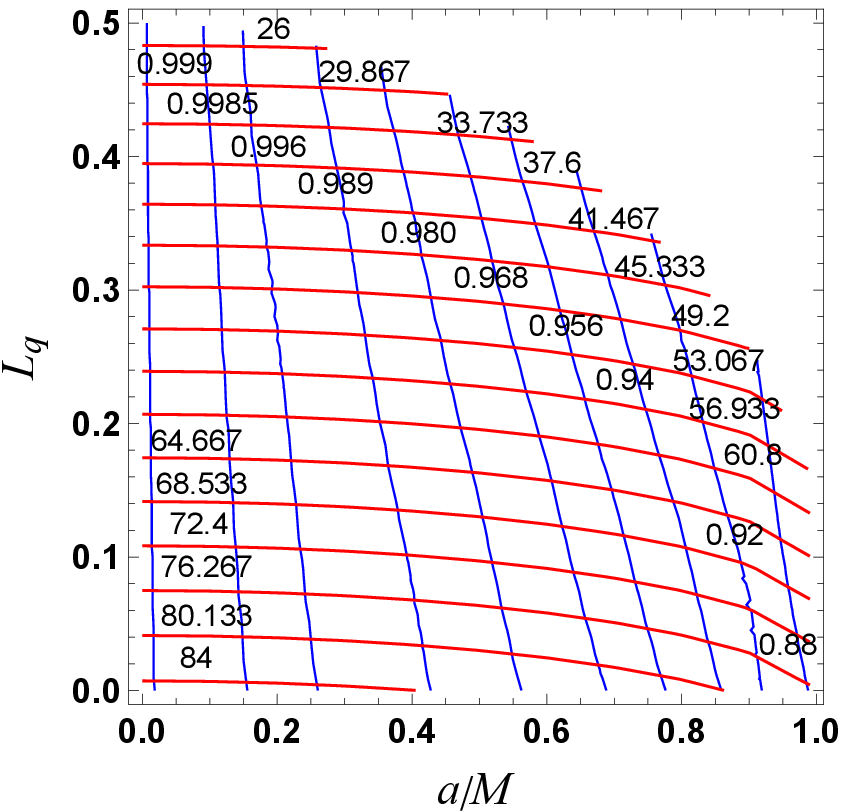}
\end{tabular}
\end{center}
	\caption{Contour plots of $A/M^2$ and $D$ in the parameter space of LIRBH-1  (left) and  LIRBH-2 (right). The red and blue curves correspond, respectively, to the contours of $A/M^2$ and $D$.}
	\label{fig:parameterEstimation}
\end{figure*}
The spherical photons (SPOs) form the BH shadow silhouette; they move on constant radii $r_p>r_+$, hitting the observer plane asymptotically far away, and are got by solving  
 \citep{Chandrasekhar:1985kt}
 \begin{equation}
\mathcal{R}(r_p)=0=\mathcal{R}'(r_p) \label{vr}
\end{equation}
Following \cite{Tsukamoto:2017fxq,Kumar:2018ple,Liu:2020ola,Brahma:2020eos,Kumar:2020ltt}, to reduce the degrees of freedom from three to two, we introduce dimensionless quantities, $\xi\equiv \mathcal{L}/\mathcal{E},\,\, \eta\equiv \mathcal{K}/\mathcal{E}^2,$
in Eq.~(\ref{R_potential}) -- which uniquely determine each light path -- and solve Eq.~(\ref{vr}) to get the critical impact parameters ($\xi_c, \eta_c$) for the SPOs,
\begin{widetext}
\begin{align}
\xi_c=&\frac{a^2-\frac{2 \Delta (r) \left(\omega'(r)+2 r\right)}{\Delta '(r)}+\omega(r)+r^2}{a},\nonumber\\
\eta_c=&-\frac{4 \Delta (r) \left(\omega'(r)+2 r\right) \left(a^2 \omega'(r)+2 a^2 r+\left(\omega(r)+r^2\right) \Delta '(r)\right)-4 \Delta (r)^2 \left(\omega'(r)+2 r\right)^2-\left(\omega(r)+r^2\right)^2 \Delta '(r)^2}{a^2 \Delta '(r)^2},\label{impactparameter}
\end{align}
\end{widetext}
where the prime symbol stands for the derivative  with respect to $r$. It turns out that the photons with $\eta_c=0$ are confined to equatorial circular trajectories, whereas $\eta_c>0$ leads to SPOs with constant radii $r_p^\mp$, which respectively stand for, the prograde and the retrograde photon radii satisfying $\eta_{c}^k=0$, $\xi_{c}^k(r_p^\mp)\gtrless0$ \citep{Tsukamoto:2017fxq,Kumar:2018ple,Afrin:2021imp,Afrin:2021wlj}.
\begin{figure*}[t]
\begin{center}
    \begin{tabular}{c c}
    \hspace{-0.81cm} \includegraphics[scale=0.90]{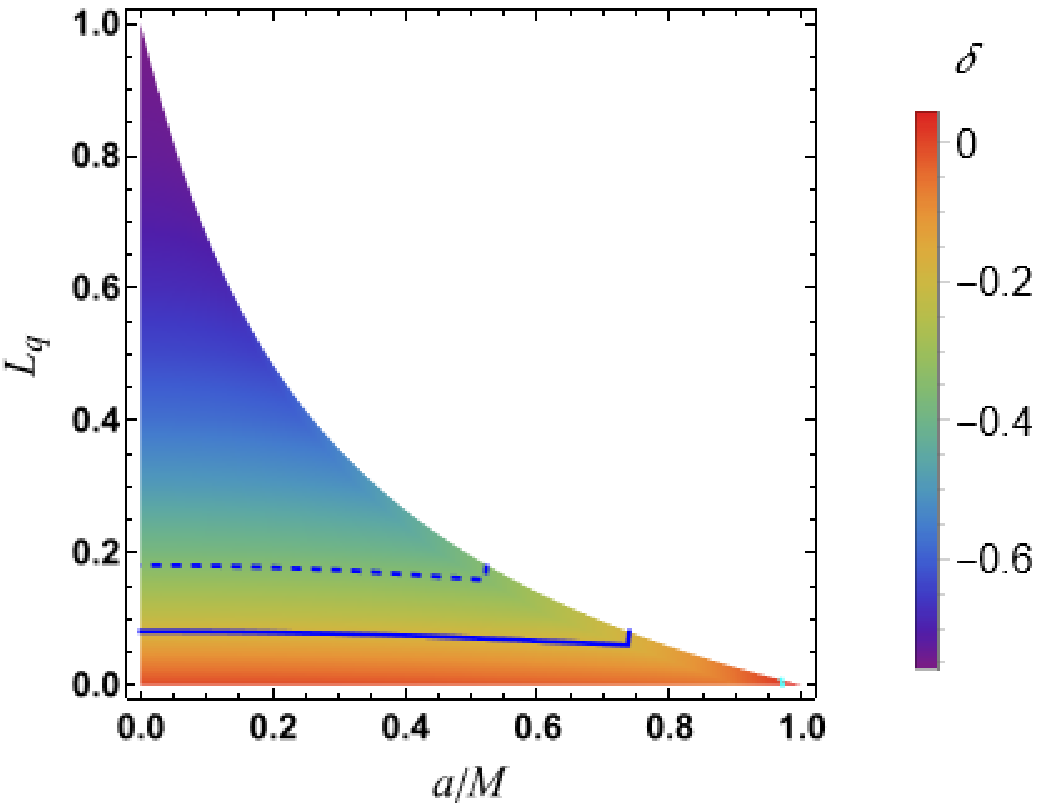}&
     \hspace{-0.48cm}\includegraphics[scale=0.90]{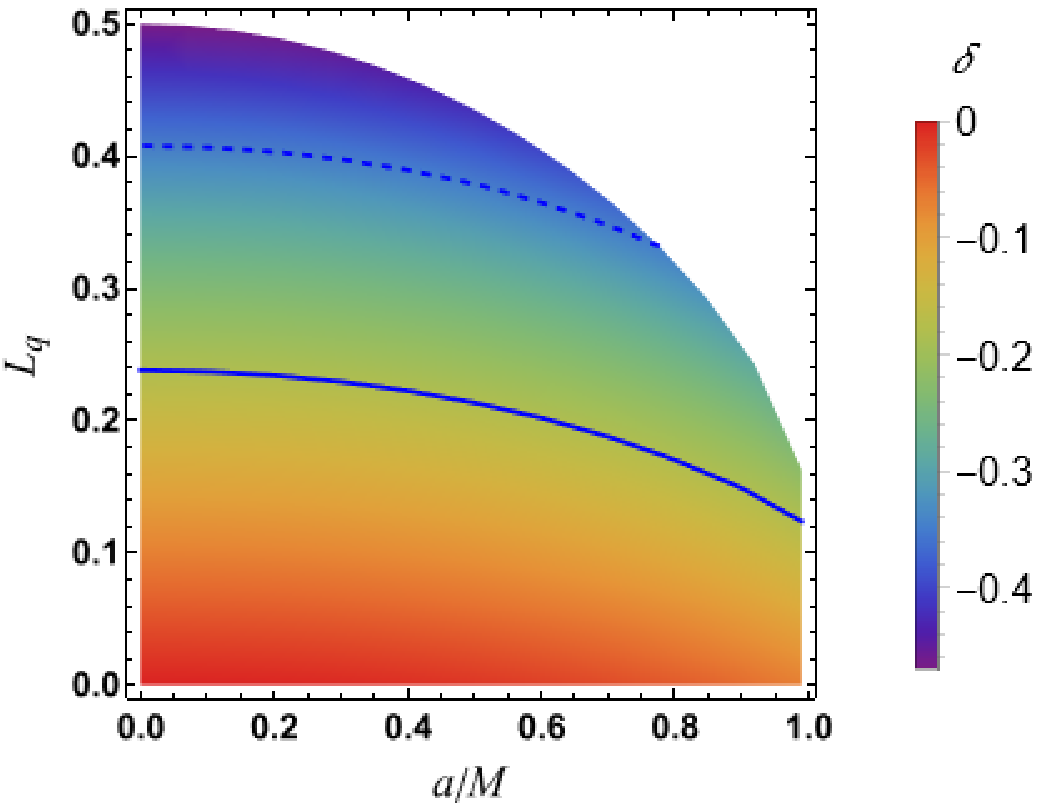}\\
   \hspace{-0.81cm} \includegraphics[scale=0.90]{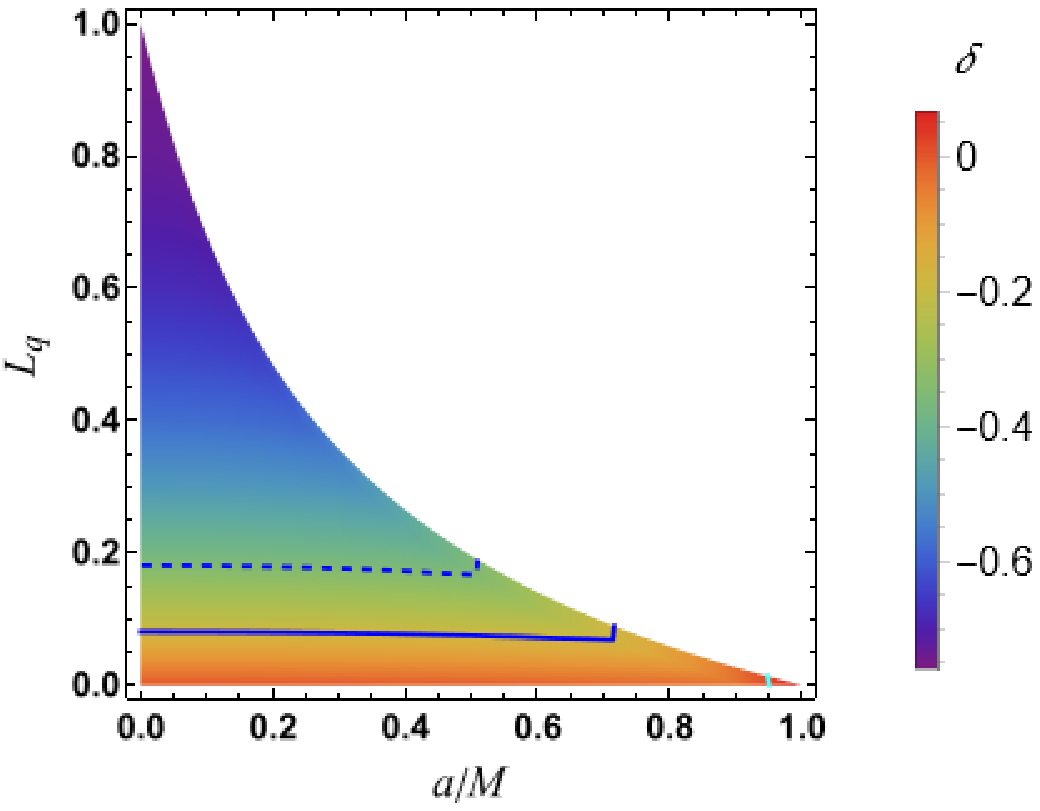}&
    \hspace{-0.48cm} \includegraphics[scale=0.90]{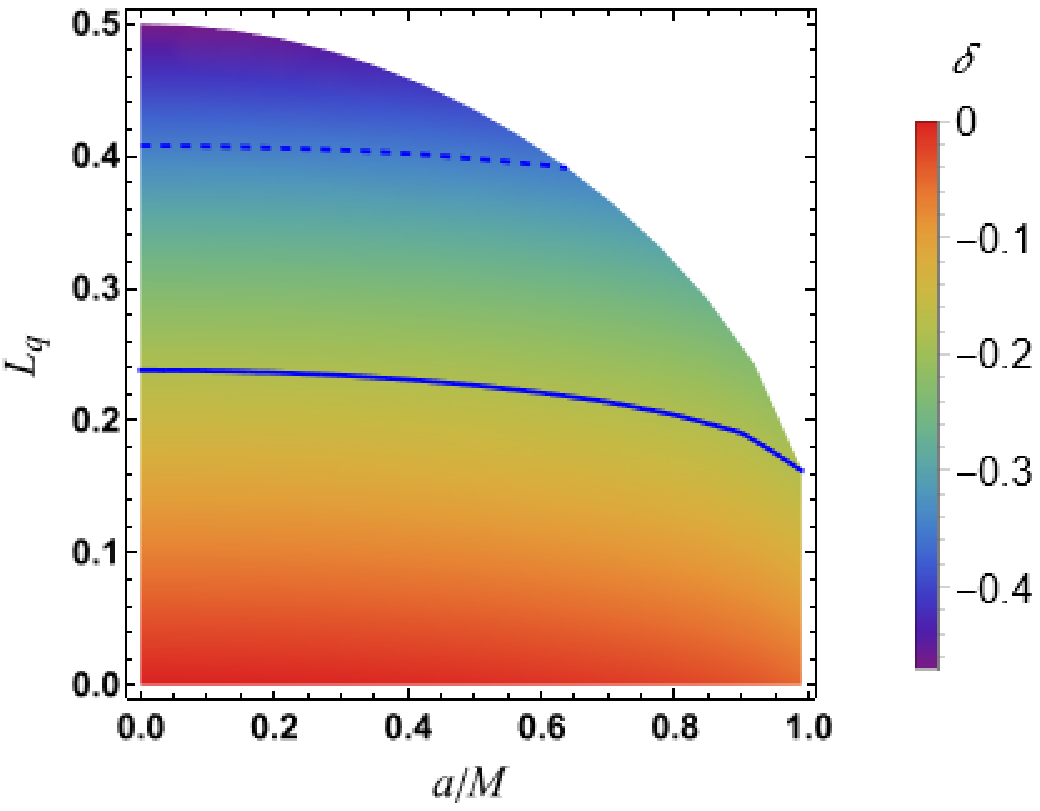}
\end{tabular}
\end{center}
	\caption{Constraints from EHT results of Schwarzschild shadow deviation $\delta$: modeling M87$^*$ as LIRBH-1  (left) and  LIRBH-2 (right) at inclinations 17\textdegree (\textit{top}) and 90\textdegree (bottom).
 The solid and dashed curves correspond, respectively, to the $1\sigma$ and $2\sigma$ bounds of the measured Schwarzschild deviation $\delta=-0.01\pm0.17$ of M87$^*$, as reported by the EHT observations.}
	\label{fig:shadowDiameter_M87}
\end{figure*}
\begin{figure*}[t]
\begin{center}
    \begin{tabular}{c c}
    \hspace{-0.7cm} \includegraphics[scale=0.90]{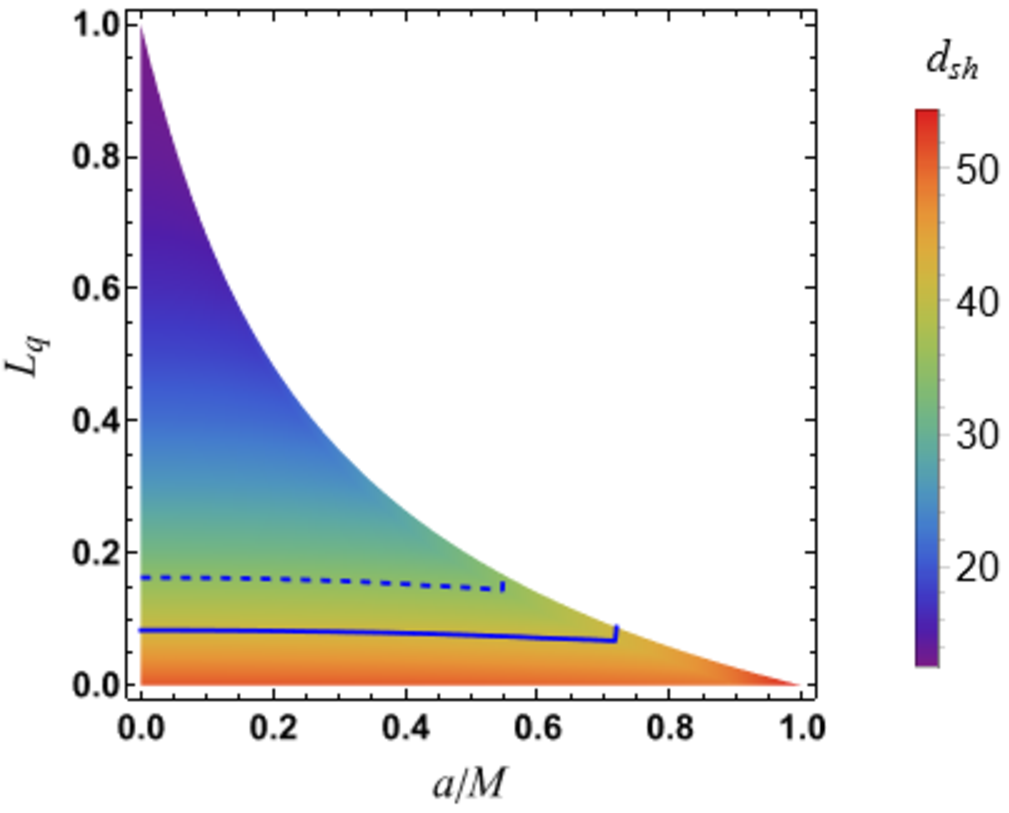}&
     \hspace{-0.4cm}\includegraphics[scale=0.90]{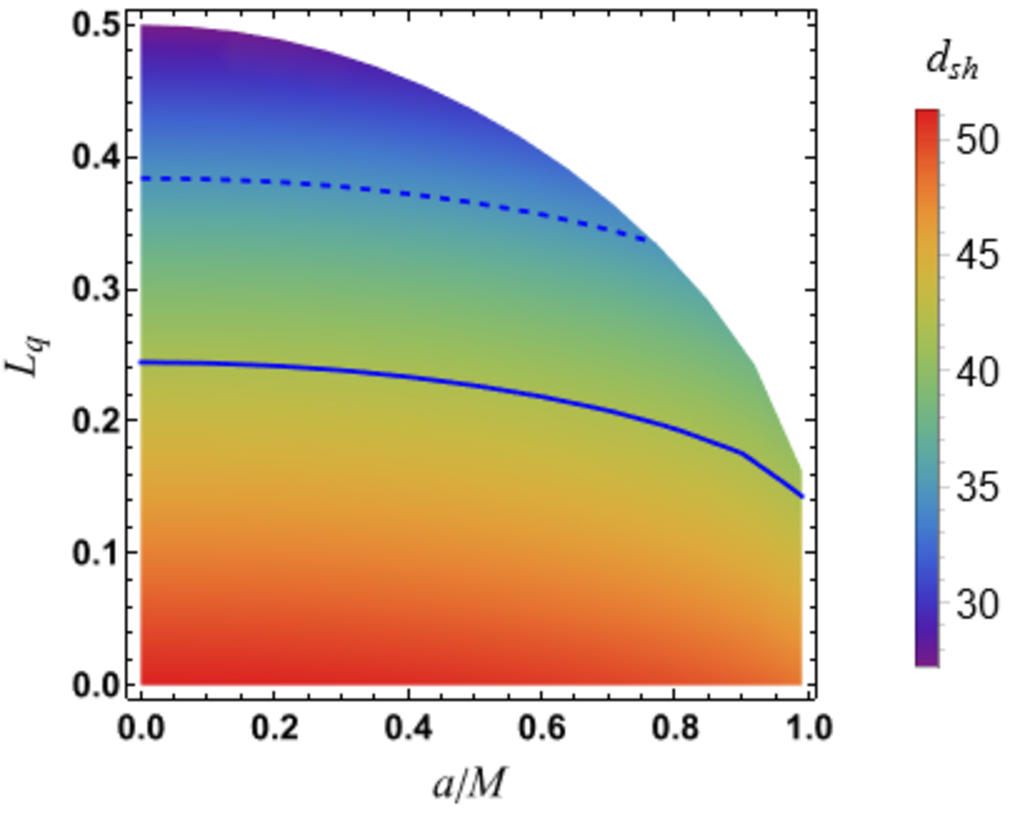}\\
   \hspace{-0.7cm} \includegraphics[scale=0.90]{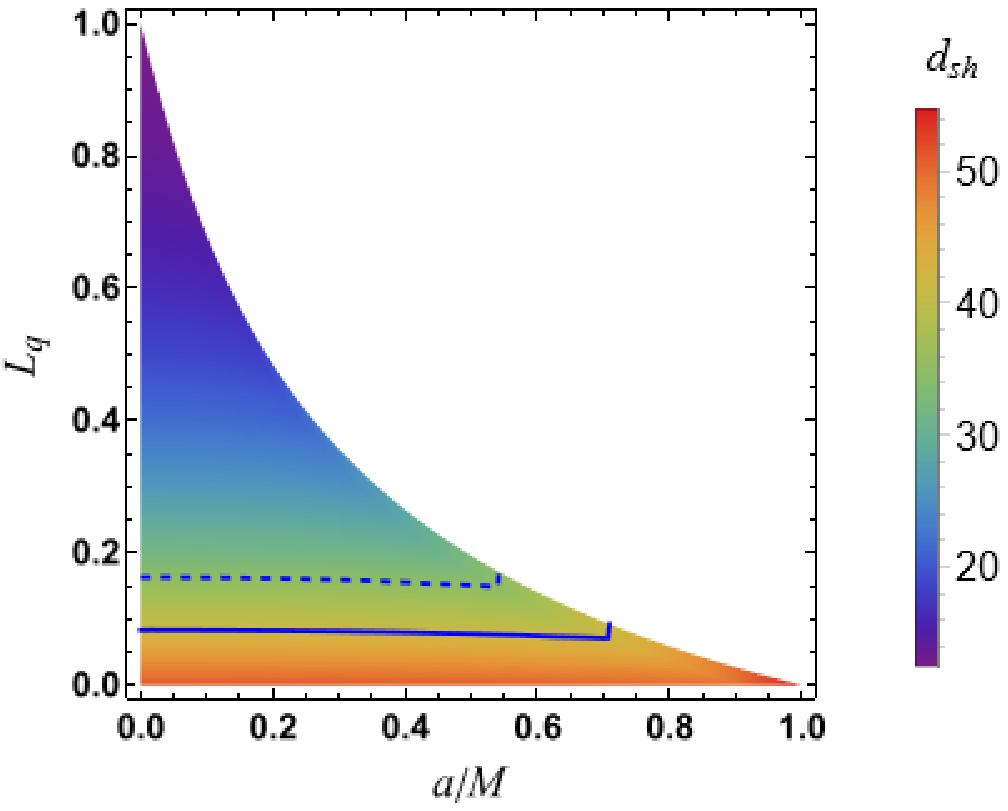}&
     \hspace{-0.4cm}\includegraphics[scale=0.90]{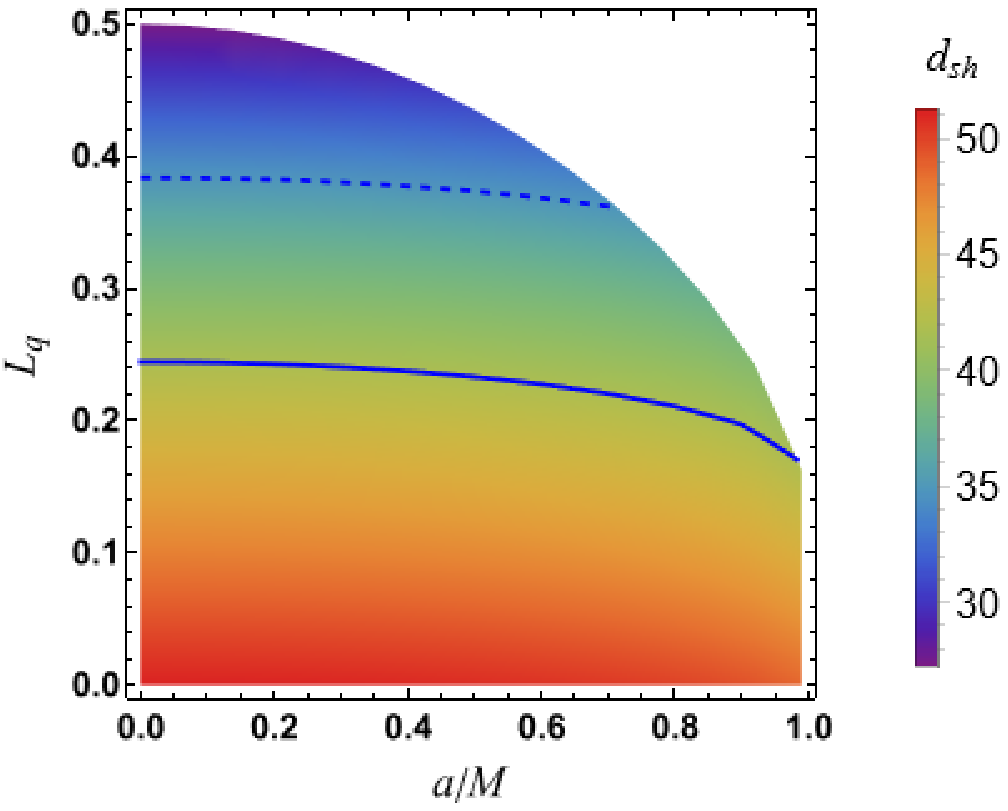}
\end{tabular}
\end{center}
	\caption{Constraints from EHT results of angular shadow diameter  $d_{sh}$: modeling Sgr A$^*$ as LIRBH-1  (left) and  LIRBH-2 (right) at inclinations 50\textdegree (\textit{top}) and 90\textdegree (bottom). The solid and dashed curves correspond, respectively, to the $1\sigma$ and $2\sigma$ bounds of the measured shadow diameter, $d_{sh}= 48.7 \pm 7\,\mu$as of Sgr A$^*$, as reported by the EHT observations.}
	\label{fig:shadowDiameter}
\end{figure*}

The gravitationally lensed image of the photon sphere around the BH yields the apparent BH shadow. For an asymptotically faraway observer ($r_0\to\infty$), making an inclination angle $\theta_o$ with the spin axis, the BH shadow is a dark region in the celestial plane outlined by a bright ring \citep{Johannsen:2015mdd,Johnson:2019ljv} with coordinates \citep{Afrin:2021wlj,Kumar:2020yem,Bardeen:1973tla}
\begin{equation}
\{X,Y\}=\{-\xi_{c}\csc\theta_0,\, \pm\sqrt{\eta_{c}+a^2\cos^2\theta_0-\xi_{c}^2\cot^2\theta_0}\}\,.\label{pt}
\end{equation} 
The shadow coordinates $\{X,Y\}$ for the LQG BHs (\ref{metric}), cast in Kerr-like form, have the same functional form as of the Kerr case \citep{Tsukamoto:2017fxq,Kumar:2018ple,Kumar:2020yem,Kumar:2020ltt,Afrin:2021imp,Afrin:2021ggx,Afrin:2021wlj,Ghosh:2022kit}; this simplifies the shadow analysis substantially.

The shadow silhouette is constructed by plotting ($X$,$Y$) in parametric form as a function of $r_p$. From Figure~\ref{shadow_Figure}, it turns out that the shadows become smaller and get more distinctly distorted with an increase in the quantum effects, $L_q$; this shows that the LQG parameter, which is expected to have significance only at the Planck scale, in reality has a non-negligible, rather profound effect on observable features like that of the shadow shape and size. We exploit this visible effect to see whether these imprints of the $L_q$ in the shadow can be exploited to extract and constrain the parameter $L_q$. Additionally, there is a horizontal shift in the shadow center along the $X$-axis, with increase $a$, because of the frame-dragging effect.
\begin{table}[t]
\caption{Estimated Parameters of the Two LGQ Model BHs}
      \centering
\begin{tabular}{c|cc|cc}
\hline
\multirow{2}{*}{Model} & \multicolumn{2}{c|}{Shadow Observables}               & \multicolumn{2}{c}{Estimated Parameters}              \\ \cline{2-5} 
                           & \multicolumn{1}{c|}{$A/M^2$} & \multicolumn{1}{c|}{$D$} & \multicolumn{1}{c|}{$L_q$} & \multicolumn{1}{c}{$a/M$} \\ \hline
LIRBH-1 &
  \multicolumn{1}{l|}{\begin{tabular}[c]{@{}l@{}}\hspace{-0.5cm}6\\ \hspace{-0.5cm}8\\ \hspace{-0.6cm}20\\ \hspace{-0.6cm}45\\ \hspace{-0.6cm}80\end{tabular}} &
  \begin{tabular}[c]{@{}l@{}}\hspace{-0.9cm}0.999\\ \hspace{-0.9cm}0.996\\ \hspace{-0.9cm}0.980\\ \hspace{-0.9cm}0.956\\ \hspace{-0.9cm}0.940\end{tabular} &
  \multicolumn{1}{l|}{\begin{tabular}[c]{@{}l@{}}\hspace{-0.8cm}0.8020\\ \hspace{-0.8cm}0.6393\\ \hspace{-0.8cm}0.1844\\ \hspace{-0.8cm}0.1240\\ \hspace{-0.8cm}0.0023\end{tabular}} &
  \begin{tabular}[c]{@{}l@{}}\hspace{-0.8cm}0.0049\\ \hspace{-0.8cm}0.0319\\ \hspace{-0.8cm}0.3226\\ \hspace{-0.8cm}0.4751\\ \hspace{-0.8cm}0.8520\end{tabular} \\ \hline
LIRBH-2 &
  \multicolumn{1}{l|}{\begin{tabular}[c]{@{}l@{}}\hspace{-0.6cm}26\\ \hspace{-0.8cm}45.33\\ \hspace{-0.8cm}60.80\\ \hspace{-0.8cm}68.53\\ \hspace{-0.8cm}76.27\end{tabular}} &
  \begin{tabular}[c]{@{}l@{}}\hspace{-0.9cm}0.999\\ \hspace{-0.9cm}0.996\\ \hspace{-0.9cm}0.980\\ \hspace{-0.9cm}0.956\\ \hspace{-0.9cm}0.920\end{tabular} &
  \multicolumn{1}{l|}{\begin{tabular}[c]{@{}l@{}}\hspace{-0.8cm}0.4816\\ \hspace{-0.8cm}0.3323\\ \hspace{-0.8cm}0.1991\\ \hspace{-0.8cm}0.1174\\ \hspace{-0.8cm}0.0276\end{tabular}} &
  \begin{tabular}[c]{@{}l@{}}\hspace{-0.8cm}0.0075\\ \hspace{-0.8cm}0.1873\\ \hspace{-0.8cm}0.4710\\ \hspace{-0.8cm}0.7107\\ \hspace{-0.8cm}0.9105\end{tabular} \\ \hline
\end{tabular}
\label{parameter_estimation_Table}
\end{table}
\begin{figure*}[t]
\begin{center}
    \begin{tabular}{c c}
   \hspace{-0.81cm}
   \includegraphics[scale=0.90]{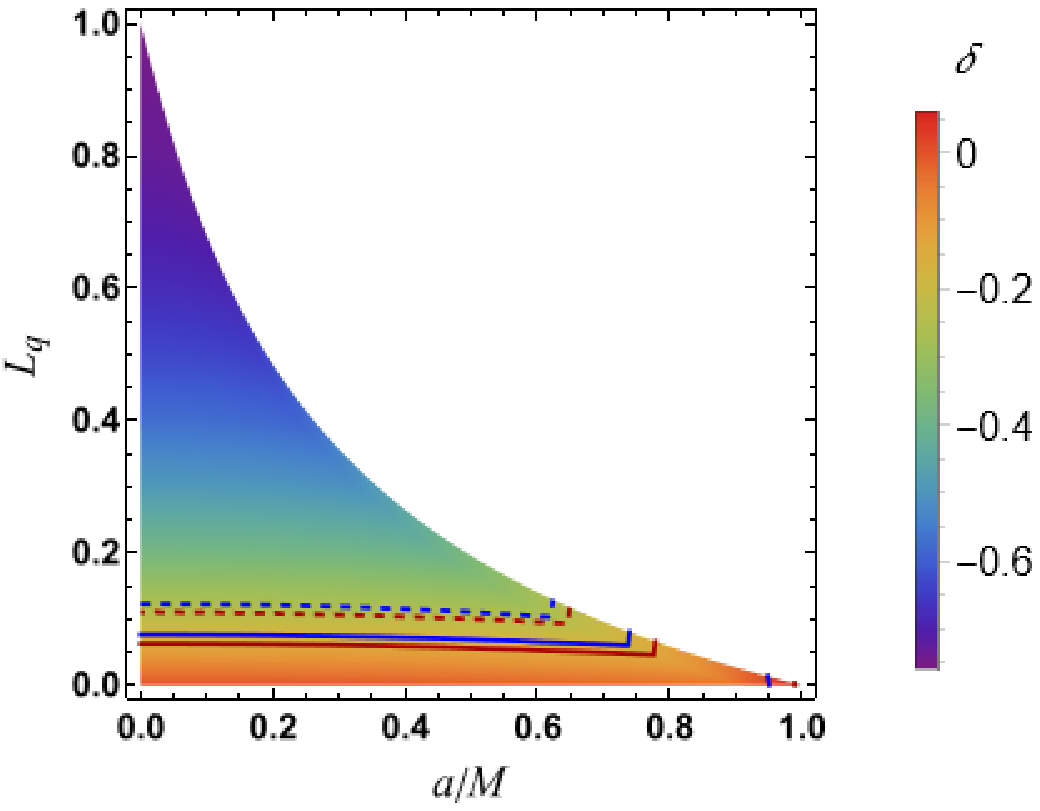}&
     \hspace{-0.48cm}\includegraphics[scale=0.90]{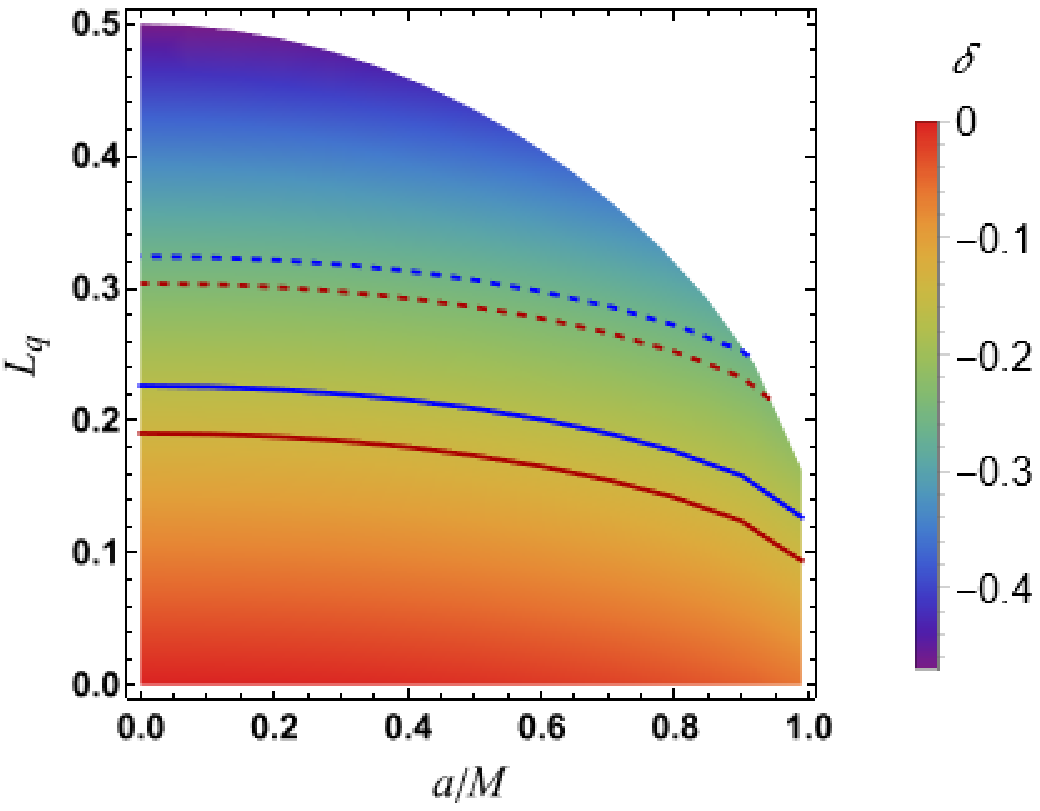}\\
     \hspace{-0.81cm}\includegraphics[scale=0.90]{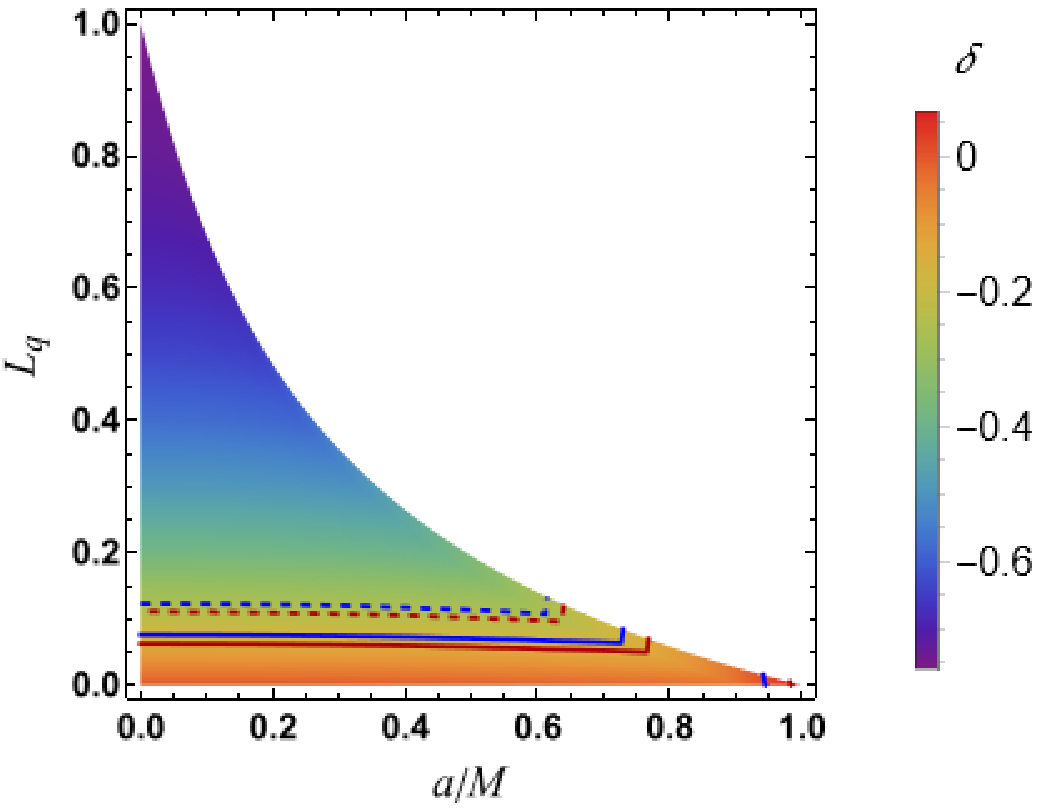}&
     \hspace{-0.48cm}\includegraphics[scale=0.90]{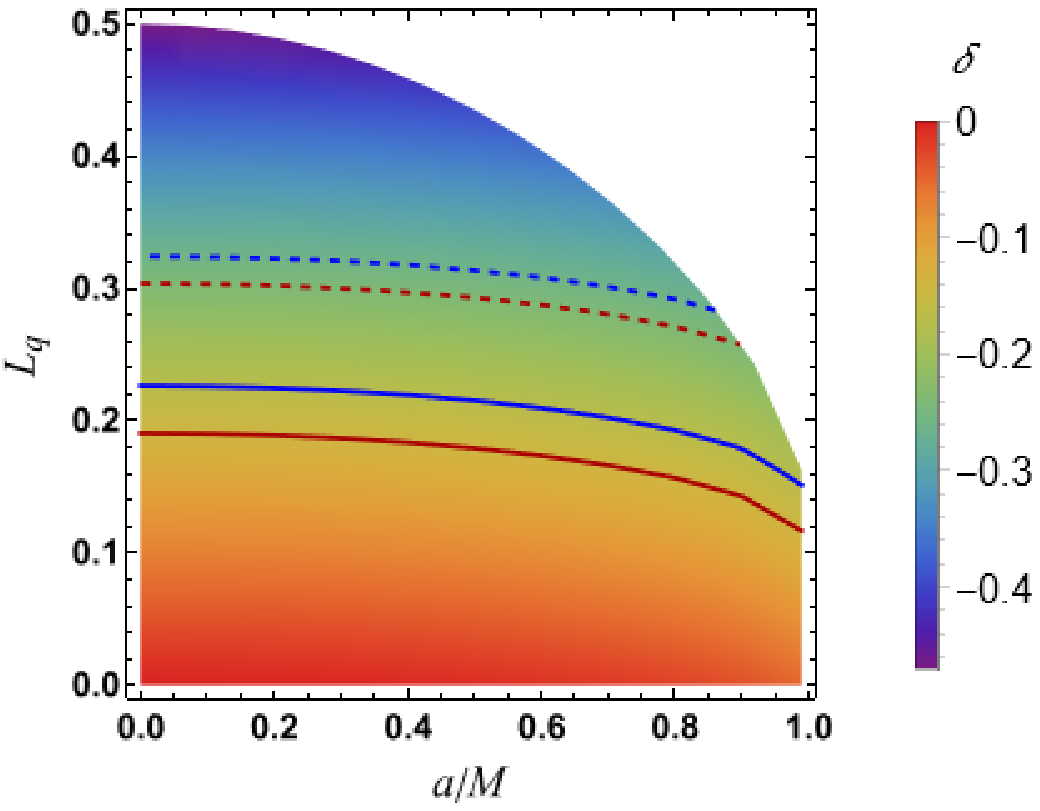}
\end{tabular}
\end{center}
	\caption{Constraints from EHT results of Schwarzschild shadow deviation $\delta$: modeling Sgr A$^*$ as LIRBH-1  (left) and  LIRBH-2 (right) at inclinations 50\textdegree (\textit{top}) and 90\textdegree (bottom). The blue and red solid contours correspond, respectively, to the $1\sigma$ bounds of the measured Schwarzschild deviation $\delta= -0.08^{+0.09}_{-0.09}~\text{(VLTI)},-0.04^{+0.09}_{-0.10}~\text{(Keck)}$ of Sgr A$^*$, as reported by the EHT observations. The dashed lines correspond to the respective $2\sigma$ bounds.}
	\label{fig:schwazrschildDeviation}
\end{figure*}
\section{Parameter estimation}\label{sect:estimation}
The BH shadow, which encodes the nature of background spacetime in its characteristic shape and size (see Figure~\ref{shadow_Figure}), can serve as a tool to test the underlying theory of gravity and to constrain the deviation parameters~\citep{Cunha:2019ikd,Banerjee:2019nnj,Allahyari:2019jqz,Yan:2019hxx,Vagnozzi:2020quf,Khodadi:2020jij,Jusufi:2020wmp,Jusufi:2021fek,Okyay:2021nnh,Roy:2021uye,Chen:2022nbb,Pantig:2022ely,Khodadi:2022pqh,Odintsov:2022umu,Oikonomou:2022tjm}. We aim to get more information about the LIRBHs, one of the most important steps in which is to extract the parameter $L_q$ observationally, which has not been done yet in the framework of LQG. Thus, we outline a simple method of BH parameter estimation using the shadow observables -- shadow area $A$ and oblateness $D$ -- which is robust in the sense that it can be employed to a haphazard shadow shape utilizing minimal symmetry \citep{Kumar:2018ple}. To estimate the LQG parameter $L_q$ besides the BH spin $a$, associated with the LIRBHs under consideration, we define the area enclosed within the shadow silhouette as \citep{Abdujabbarov:2015xqa,Kumar:2018ple}
\begin{eqnarray}
A&=&2\int{Y(r_p) dX(r_p)}\nonumber\\
&=&2\int_{r_p^{-}}^{r_p^+}\left( Y(r_p) \frac{dX(r_p)}{dr_p}\right)dr_p,\label{Area}
\end{eqnarray} 
where $r_p^{\mp}$ are, respectively, the prograde and retrograde SPO radii obtained as the smallest and largest real roots of :  $\eta_{c}=0$, $\xi_{c}(r_p^\mp)\gtrless0$, outside the event horizon \citep{Teo:2020sey}.
Next, we quantify the deformation in shadow shape -- induced by $L_q$ and $a$ -- from a perfect circle, with the shadow oblateness ($D$) observable, which can be written as \citep{Tsupko:2017rdo,Kumar:2018ple},
\begin{eqnarray}
D=\frac{X_r-X_l}{Y_t-Y_b}\label{Oblateness}
\end{eqnarray}
where the subscripts $l$, $r$, $t$ and $b$ stand for the left and right ends of the shadow silhouette, where $Y(r_p)=0$, for positive $a$, and the top and bottom points, where $Y'(r_p)=0$, respectively \citep{Hioki:2009na}. For spherically symmetric BH it is straightforward to understand that $D=1$. However, for the rotating BHs with extra deviation parameters, characteristically $D\neq1$; in the Kerr BH $1\leq D\leq \sqrt{3}/2$ \citep{Tsupko:2017rdo,Kumar:2018ple,Afrin:2021imp}. Note also that $D$ is closely related to other measures of oblatness, e.g.\ the deviation from circularity studied in~\cite{Bambi:2019tjh} (see also~\cite{Vagnozzi:2019apd}).

The parameters of the background theory of LQG -- treated as intrinsic parameters of the model BHs -- can be extracted from the shadow observables \citep{Hioki:2009na, Kumar:2018ple,Afrin:2021imp,Afrin:2021wlj} if the extrinsic parameters, namely $\theta_o$ and distance of the BH $d$ from the observer, can be measured independently. The one-to-one correspondence between the shadow characteristics, i.e., $A$ and $D$, and the BH parameters $L_q$ and $a$ is evident from Figure~\ref{shadow_Figure} \citep{Kumar:2018ple,Afrin:2021wlj}. The maximum deformation in shadow shape, a deviation from a perfect circle, is observed only at a high inclination angle, and thus we fix $\theta_0=90$\textdegree\, for estimating the parameters. The constant contours of $A/M^2$ and $D$ are degenerate in $L_q$ and $a/M$ individually, but the degeneracy is broken if they are considered together. The contours of $A/M^2$ and $D$ for any $L_q$ and $a/M$ are found to intersect at unique points (see Figure~\ref{fig:parameterEstimation}) and the coordinates of the intersections uniquely determine the two BH parameters $L_q$ and $a/M$. We tabulate selected estimated parameters of the LIRBHs in Table~\ref{parameter_estimation_Table}.
\section{Constraining with EHT observations}\label{sect:constraining}
\begin{table*}[t]
\caption{Constraints on the LQG Parameter $L_q$ for the Two Models, Set by the EHT Results.}
      \centering
\begin{tabular}{l|ll|ll}
\hline
\multirow{2}{*}{Model} & \multicolumn{2}{l|}{\quad\quad\quad\quad\quad\quad M87*}                                        & \multicolumn{2}{l}{\quad\quad\quad\quad\quad\quad Sgr A*}                                       \\ \cline{2-5} 
                       & \multicolumn{1}{l|}{\quad\quad\quad 1$\sigma$}            & \quad\quad\quad 2$\sigma$            & \multicolumn{1}{l|}{\quad\quad\quad1$\sigma$}            & \quad\quad\quad 2$\sigma$            \\ \hline
LIRBH-1                & \multicolumn{1}{l|}{$L_q\in [0, 0.0643)$} & $L_q\in [0, 0.1643)$ & \multicolumn{1}{l|}{$L_q\in [0, 0.0423)$} & $L_q\in [0, 0.0902)$ \\ \hline
LIRBH-2                & \multicolumn{1}{l|}{$L_q\in [0, 0.1253)$} & $L_q\in [0, 0.3350)$ & \multicolumn{1}{l|}{$L_q\in [0, 0.0821)$} & $L_q\in [0, 0.1834)$ \\ \hline
\end{tabular}
\vspace{1ex}

     {\raggedright Note. The available observable $d_{sh}$ is employed to constrain $L_q$ from EHT results of M87*, while the observables $d_{sh}$ and $\delta$ of Sgr A* together constrain $L_q$. \par}
\label{Constraints_Table}
\end{table*}
We set up now a framework for directly translating the bounds from the EHT results of M87* and Sgr A* to the LIRBHs, utilizing the characteristic shadow observables -- angular shadow diameter ($d_{sh}$) and Schwarzschild shadow deviation ($\delta$) -- that capture the details of the background theory of gravity.
The angular diameter of a shadow, for an observer at distance $d$ from the BH, is defined as
\citep{Kumar:2020owy,Ghosh:2020spb,Afrin:2021imp,Afrin:2021wlj}
\begin{eqnarray}
d_{sh}=2\frac{R_a}{d}\;,\;R_a=\sqrt{A/\pi},\label{angularDiameterEq}
\end{eqnarray}  
where $R_a$ is the areal shadow radius. Apart from distance $d$, we note that $d_{sh}$ implicitly depends on the mass $M$ and parameter $L_q$  of the BHs (\ref{metric}) besides the observation angle $\theta_o$.
Using EHT-considered mass and distance of M87* and Sgr A*, we calculate the angular diameter of the shadows for the two LIRBHs in question. 

The EHT images of both  M87* and Sgr A* exhibit a thick luminous ring of emission, with diameters 42 $\pm$ 3 $\mu$as and 51.8 $\pm$ 2.3 $\mu$as, respectively -- consistent with the expectations from a central supermassive BH \citep{EventHorizonTelescope:2019dse,EventHorizonTelescope:2022xnr} -- surrounding a brightness depression, namely the BH shadow \citep{EventHorizonTelescope:2019dse,EventHorizonTelescope:2022xqj}. To
 quantify the difference between the model shadow diameter ($\Tilde{d}_{metric}$) and the Schwarzschild shadow diameter $6\sqrt{3}M$ , we introduce the Schwarzschild shadow deviation ($\delta$) given by \citep{EventHorizonTelescope:2022xnr,EventHorizonTelescope:2022xqj},
\begin{equation}\label{SchwarzschildShadowDiameter}
    \delta=\frac{\Tilde{d}_{metric}}{6\sqrt{3}}-1.
\end{equation}
Here $\Tilde{d}_{metric}=2R_a$ where $R_a$ is given by Eq.~(\ref{angularDiameterEq}). In the case of Kerr BHs, $\delta\in[-0.075,0]$ \citep{EventHorizonTelescope:2022xqj}, with the variations $a\in[0,M]$ and $\theta_0\in[0,\pi/2]$, and thus for any BH to cast shadows consistent with those of Kerr BHs' should be within this range. Thus, theories of gravity predicting shadows smaller than ($\delta<-0.075$) and larger than ($\delta>0$) Kerr BHs can, aided by the $\delta$ observable, be tested.
Interestingly, the LQG BH models we consider cast shadows that are distinctly smaller and more distorted than the corresponding Kerr BH shadows and can thus be tested and constrained with EHT results. With the mass and distance of M87* and Sgr A* as considered by EHT \citep{EventHorizonTelescope:2019dse,EventHorizonTelescope:2022xnr,EventHorizonTelescope:2022xqj}, we calculate the Schwarzschild deviation of the shadows cast by the LIRBHs.

There are some caveats to the present analysis related to the observational appearance of M87* and Sgr A*. They include uncertainties induced by the different telescopes in the sparse array and more fundamental ones due to the still uncertain radiative and accretion physics \citep{Gralla:2020pra} that obfuscate the actual predictions of the EHT; any analytical testing of the theories of gravity would undoubtedly be subject to this \citep{Afrin:2021wlj}. Despite these uncertainties, the theoretical analysis, utilizing the EHT observational bounds, can serve as an initial probe of LQG, which would call for further scrutiny with future, more precise observations. 
\paragraph{M87* bounds.} 
Using an extensive library of ray-traced GR magnetohydrodynamic (GRMHD) simulations of BHs, the EHT has inferred a central compact mass of M87*, $M_{M87^*}= 6.5\times 10^9 M_\odot$ which is consistent with the previous stellar dynamical measurements, and distance of $d_{M87^*}=16.8$ Mpc from Earth \citep{EventHorizonTelescope:2019dse,EventHorizonTelescope:2019pgp,EventHorizonTelescope:2019ggy}. For simplicity, we do not consider in our analysis the possible uncertainties in the mass and distance measurements of the target BH, as the EHT results already take into consideration the various uncertainties to get the bounds on the observables. Although the characteristic features and dimensions of the observed image of  M87* are consistent with the expected appearance of Kerr BHs in GR, still the current uncertainties in the measurement of spin, inclination angle, and the relative deviation of quadrupole moments do not entirely rule out Kerr-like BHs in modified gravity theories \citep{EventHorizonTelescope:2019dse,EventHorizonTelescope:2019pgp,EventHorizonTelescope:2019ggy,Cardoso:2019rvt}, including those in LQG. But to be consistent with the dimensional expectations of the corresponding Kerr BH's shadow at a spin $a$, the results of the 2019 EHT drive can put constraints on the LQG parameter $L_q$, as we shall explore here. Previously, the parameter space of the LIRBH-2 has been constrained with the shadow diameter of M87* at $\theta_0=17$\textdegree, and the wormhole region III has been ruled out \citep{Brahma:2020eos}. We get the numerical value of the constraints with M87* so as to be able to compare with the constraints obtained with results of Sgr A*,  at both $\theta_0=90$\textdegree and 17\textdegree. Meanwhile, we also reaffirm the results of the earlier work  \citep{Brahma:2020eos}. 
Calibrating the size of the shadow of M87*, with the ring diameter that the EHT has measured, yields the $1\sigma$ bound $\delta=-0.01\pm0.17$ \citep{EventHorizonTelescope:2019ggy,EventHorizonTelescope:2020qrl,EventHorizonTelescope:2021dqv}. The constant $1\sigma$ and $2\sigma$ contours
delimit a finite parameter space of the LIRBHs, as seen from Figure~\ref{fig:shadowDiameter_M87}, and upper limits can be placed on $L_q$. For LIRBH-1, $L_q\in [0, 0.1687)$ 
within $2\sigma$, $L_q\in [0, 0.0686)$ within $1\sigma$ confidence levels at $\theta_0=90$\textdegree and $L_q\in [0, 0.1643)$ 
within $2\sigma$, and $L_q\in [0, 0.0643)$ within $1\sigma$  confidence levels at $\theta_0=17$\textdegree. For LIRBH-2, $L_q\in [0, 0.3904)$ 
within $2\sigma$, $L_q\in [0, 0.1633)$ within $1\sigma$ confidence levels at $\theta_0=90$\textdegree and $L_q\in [0, 0.3350)$ 
within $2\sigma$, and $L_q\in [0, 0.1253)$ within $1\sigma$ confidence levels at $\theta_0=17$\textdegree. Thus, we infer the constraints $L_q\in [0, 0.0643)$ for LIRBH-1 and $L_q\in [0, 0.1253)$ for LIRBH-2 from the EHT results of M87$^*$.
\paragraph{Sgr A* bounds.} Though the shadow of the M87* BH provides a ground to constrain parameter $L_q $ of the LQG theories \citep{Brahma:2020eos}, the observations of Sgr A* would offer independent tests in a much higher curvature regime, as a consequence of $\mathcal{O}(M_{Sgr A^*})\sim10^6 M_\odot$ being smaller than $\mathcal{O}(M_{M87^*})\sim10^9 M_\odot$ by several order; thus, we can leverage the varied range of conditions that can be probed with the two different target BHs \citep{EventHorizonTelescope:2022xqj,Vagnozzi:2022moj}. Besides, for Sgr A*, the ratios between the mass $M_{SgrA^*} = 4.0 \times 10^6 M_\odot $ and distance from Earth $d_{SgrA^*}=8 kpc$ \citep{EventHorizonTelescope:2022xnr,EventHorizonTelescope:2022xqj} could be used as priors from independent observations stellar dynamic observations of the S0-2 star's orbits by Keck telescopes and Very Large Telescope Interferometer (VLTI) \citep{Do:2019txf,GRAVITY2019,GRAVITY2021,GRAVITY2022,EventHorizonTelescope:2022xqj} and the predicted size of the Kerr shadow can be directly compared to the observations \citep{EventHorizonTelescope:2022xqj}.

From the observed image of Sgr A*, the EHT, besides obtaining the diameter of the bright emission ring, has also measured the shadow diameter $d_{sh}= 48.7 \pm 7\,\mu$as and Schwarzschild shadow deviation $\delta = -0.08^{+0.09}_{-0.09}~\text{(VLTI)},-0.04^{+0.09}_{-0.10}~\text{(Keck)}$ at the $1\sigma$ confidence level; the image is consistent with the shadow of the Kerr BH in GR \citep{EventHorizonTelescope:2022xnr,EventHorizonTelescope:2022xqj}. 
We model Sgr A* as the LIRBHs and impose the EHT-inferred bounds on $d_{sh}$ (see Figure~\ref{fig:shadowDiameter}) to find the constraints on the parameters $L_q$. We obtain the following constraints: (i) for LIRBH-1, $L_q\in [0, 0.1471)$ 
within $2\sigma$, $L_q\in [0, 0.0761)$ within $1\sigma$ confidence levels at $\theta_0=90$\textdegree, $L_q\in [0, 0.1439)$ 
within $2\sigma$, $L_q\in [0, 0. 0690)$ within $1\sigma$  confidence levels at $\theta_0=50$\textdegree and $L_q\in [0, 0.1415)$ 
within $2\sigma$, $L_q\in [0, 0.0683)$ within $1\sigma$  confidence levels at $\theta_0=0$\textdegree and (ii) for LIRBH-2, $L_q\in [0, 0.3627)$ 
within $2\sigma$, $L_q\in [0, 0.1707)$ within $1\sigma$ confidence levels at $\theta_0=90$\textdegree, $L_q\in [0, 0.3374)$ 
within $2\sigma$, $L_q\in [0, 0. 1417)$ within $1\sigma$  confidence levels at $\theta_0=50$\textdegree and $L_q\in [0, 0.2832)$ 
within $2\sigma$, $L_q\in [0, 0.1260)$ within $1\sigma$  confidence levels at $\theta_0=0$\textdegree.
Next, we impose the bounds on $\delta$  (see Figure~\ref{fig:schwazrschildDeviation}) and get the EHT-consistent parameter ranges: (i) for LIRBH-1, $L_q\in [0, 0.0968)$ 
within $2\sigma$, $L_q\in [0, 0.0492)$ within $1\sigma$ confidence levels at $\theta_0=90$\textdegree, $L_q\in [0, 0.0913)$ 
within $2\sigma$, $L_q\in [0, 0. 0437)$ within $1\sigma$  confidence levels at $\theta_0=50$\textdegree and $L_q\in [0, 0.0902)$ 
within $2\sigma$, $L_q\in [0, 0.0423)$ within $1\sigma$  confidence levels at $\theta_0=0$\textdegree; and (ii) for LIRBH-2, $L_q\in [0, 0.2566)$ 
within $2\sigma$, $L_q\in [0, 0.1170)$ within $1\sigma$ confidence levels at $\theta_0=90$\textdegree, $L_q\in [0, 0.2145)$ 
within $2\sigma$, $L_q\in [0, 0. 0955)$ within $1\sigma$  confidence levels at $\theta_0=50$\textdegree and $L_q\in [0, 0.1834)$ 
within $2\sigma$, $L_q\in [0, 0.0821)$ within $1\sigma$  confidence levels at $\theta_0=0$\textdegree. We note that the bounds on the two observables $d_{sh}$ and $\delta$ are very similar and thus comparable, as is clear from the obtained limits on the BH parameters; comparing all the upper limits, we infer that the upper bound on $L_q$ are  $L_q^{max}\in[0.0423, 0.0492]$ for LIRBH-1 and $L_q^{max}\in[0.0821, 0.1170]$ for LIRBH-2 as $\theta_0$ varies from $0$\textdegree to $90$\textdegree. The constraints on $L_q$ inferred from the EHT results of M87* and Sgr A* are summarized in Table~\ref{Constraints_Table}.
\section{Conclusions}\label{sect:conclusion}
The EHT collaboration has determined the expected shadow size of the supermassive BH Sgr A*, based on previous information on the mass-to-distance ratio of the BH. The results agree with the Kerr metric's prediction, and there is no evidence of any violations of the theory of GR. Sgr A* has the largest mass-to-distance ratio amongst the data for the different BHs, which makes it the optimal target for testing the no-hair theorem. 

We show that the shadows of these LIRBHs differ significantly from those of Kerr BHs with the same spin and indicate the feasibility of testing the no-hair theorem by constraining the deviation parameter $L_q$ associated with LIRBHs with  EHT results of Sgr A*. LIRBHs are modifications of the Kerr spacetime, e.g., the null geodesic structure of the spacetime gets modified---which is most important for our purposes---leading to substantial changes in the properties, which may be valuable to empirically test the no-hair theorem~\citep{Johannsen:2010ru}. 
We have considered two LIRBHs that resolve the singularity problem in GR and, in the absence of quantum effects ($L_q=0$), go over to the Kerr metric. To construct the shadow, we solve the Hamilton--Jacobi equations and find that they are still separable and yield first-order photon geodesic equations.

Interestingly, the shadow silhouettes exhibit deviations in the characteristic shape and size from the Kerr BH shadows; the shadows are smaller and more distorted as $L_q$ increases. We further investigate the possibility of degeneracy between the shadows of LIRBHs and those of the Kerr BHs by constructing shadow observables $A$, $D$, $d_{sh}$ and $\delta$ that quantify the shadow dimensions and deformations. Using observables $A$ and $D$, we follow a simple contour intersection technique to estimate the quantum parameter $L_q$ and the BH spin $a$, which accord additional information about the quantum nature of gravity. 

Further, modeling M87* as LIRBHs and imposing the observational bounds on the $d_{sh}$ observable at different inclinations, we get the EHT-consistent range of the LQG parameters: $0\leq L_q< 0.0643$ for LIRBH-1 and $0\leq L_q< 0.1253$ for LIRBH-2. Intending to probe the LQG at a different curvature scale, we impose the observational bounds of Sgr A* on two of the shadow observables, $d_{sh}$ and $\delta$, to find that the astrophysical allowed ranges of the LQG parameter become more constricted: $0\leq L_q< 0.0423$ for LIRBH-1 and $0\leq L_q< 0.0821$  for LIRBH-2. Thus, with observational results of Sgr A*, we can place more stringent bounds on both the LIRBHs than those that we get from M87*.

The singularities of the Kerr BH spacetime - a limitation of the classical theory of GR, are likely to be resolved in the LQG, e.g., LIRBHs are regular everywhere and go to the Kerr solution in the absence of quantum effects ($L_q=0$). However, the LIRBH metrics do not result from a direct loop quantization of the Kerr spacetime, but these models furnish singularity resolution of Kerr BHs; LIRBHs can capture the effective regular spacetime description of LQG and hence can be suitable candidates for astrophysical BHs.

Many interesting avenues are amenable to future work; it will be intriguing to analyze accretion models in LIRBHs. Since we find that the LQG parameter profoundly influences the shadow, it may have several astrophysical consequences, e.g., gravitational lensing. In the spirit of the no-hair theorem, one can consider a further detailed analysis of the two LIRBHs with different astronomical observations.

\section*{Acknowledgements}
M.A.\ is supported by a DST-INSPIRE Fellowship, Department of Science and Technology, Government of India. S.V.\ acknowledges a College Research Associateship at Homerton College, University of Cambridge. S.G.G.\ is supported by SERB-DST through project No.~CRG/2021/005771.
\bibliography{sample63}{}
\bibliographystyle{aasjournal}

\end{document}